\documentclass[reprint,aps,pra,amsmath,amssymb,superscriptaddress]{revtex4-2}

\usepackage{graphicx}
\usepackage{dcolumn}
\usepackage{amsmath}
\usepackage{amssymb}
\usepackage{mathtools}
\usepackage{bbm}
\usepackage{bm}
\usepackage[caption=false]{subfig}
\usepackage{placeins}
\usepackage{amsthm}
\usepackage{natbib}
\usepackage[colorlinks=true,citecolor=blue,linkcolor=blue,urlcolor=blue]{hyperref}

\newcommand{\upe}{\mathrm{e}}   
\newcommand{\upi}{\mathrm{i}}   
\newcommand{\upd}{\mathrm{d}}   

\begin{document}

\title{Multi-ensemble superradiance for distributed quantum sensing}

\author{Kang Shen}
\affiliation{College of Physical Science and Technology, Central China Normal University, Wuhan 430079, China}

\author{Xiangming Hu}
\email{xmhu@ccnu.edu.cn}
\affiliation{College of Physical Science and Technology, Central China Normal University, Wuhan 430079, China}

\author{Fei Wang}
\email{feiwang@hbut.edu.cn}
\affiliation{School of Science, Hubei University of Technology, Wuhan 430068, China}

\begin{abstract}
Multi-ensemble superradiance extends Dicke superradiance to multiple ensembles and supports dark states whose properties depend on the initial state. In the large-\(N\) limit, we derive analytical covariance matrices for these dark states, revealing inter-ensemble entanglement that enhances quantum metrology. The minimum eigenvalue, determined by the curvature of the superradiance potential, corresponds to the optimal multiparameter spin-squeezing coefficient, which is given by the \emph{Rayleigh quotient} of the spin-squeezing matrix, linking metrological sensitivity to the geometric structure of the underlying dynamics. The multiparameter squeezing coefficient provides a variational framework for optimizing metrological performance. These results enable optimal estimation of arbitrary linear combinations of multiple parameters, offering a concrete protocol for distributed quantum sensing and a promising route toward multimode quantum interferometry.
\end{abstract}
\maketitle

\section{Introduction}
Entanglement and squeezing, as key nonclassical resources for quantum-enhanced metrology, have been extensively investigated for their ability to surpass the standard quantum limit (SQL) imposed by classical resources in single-parameter estimation~\cite{Pezze2018,Abbott2016,Bao2020,Degen2017,Dutt2015,Esteve2008,Eckner2023,Braverman2019,Bornet2023,Muessel2014,Takano2009,wu2025}. However, many metrological tasks, including imaging and field sensing, require the simultaneous estimation of multiple parameters, motivating the development of multiparameter techniques, with particular emphasis on strategies exploiting mode entanglement, thereby enabling quantum-enhanced measurement precision in multiparameter estimation beyond what is achievable with independently squeezed modes.~\cite{Altenburg2017,Gessner2020,Fadel2023,Wildermuth2006,Koschorreck2011,Guo2020,Proctor2018,Ge2018,Kaubruegger2023}. In this context, distributed quantum sensing, as a paradigmatic task in multiparameter estimation, aims to estimate a global parameter from spatially distributed unknown parameters with quantum-enhanced sensitivity~\cite{Ge2018,Proctor2018,Oh2020,Zhang2021,Oh2022,Liu2024} and has attracted significant interest for applications such as quantum imaging~\cite{Brida2010,Defienne2019}, sensor networks~\cite{Eldredge2018,Gessner2020}, phase tracking~\cite{Yonezawa2012}, dark matter searches~\cite{Backes2021,Shi2023,Brady2023}, and global-scale clock synchronization~\cite{Komar2014,Malia2022}. Recent experimental~\cite{Kunkel2018,Matteo2018} and theoretical~\cite{Kajtoch2018,Jing2019,Fadel2020} studies have realized entanglement between addressable modes using split spin-squeezed ensembles generated via one-axis twisting dynamics~\cite{Kitagawa1993}, as well as split Dicke states~\cite{Karsten2018}. The resulting mode-entangled atomic ensembles have been experimentally demonstrated to enhance distributed quantum sensing~\cite{Malia2022}.

Despite these advances, many existing studies of distributed quantum sensing and multiparameter metrology rely on pre-engineered intermode correlations, which are introduced as external and static resources for the sensing protocol~\cite{Karsten2018,Jing2019,Malia2022,Matteo2018,Kunkel2018}. By contrast, collective light–matter interaction~\cite{Matthew2018,Lewis2018} provides a natural mechanism for generating strong, dynamical correlations among multiple ensembles. In particular, Dicke superradiance~\cite{Dicke1954} offers a paradigmatic platform in which collective dissipation induces robust many-body correlations and enables quantum-enhanced metrological performance. While the driven Dicke superradiance, also known as the cooperative resonance fluorescence model~\cite{Puri1979,Carmichael1980}, has been extensively studied in single-ensemble settings, revealing rich nonequilibrium phenomena such as dissipative phase transitions~\cite{Puri1979,Carmichael1980} and spin squeezing~\cite{Barberena2019,Lee2014}, the extension to multiple, spatially separated ensembles remains largely unexplored. Such a multi-ensemble generalization not only introduces additional collective degrees of freedom but also naturally gives rise to mode entanglement across ensembles, thereby providing a promising route toward intrinsically generated resources for multiparameter quantum estimation.

\begin{figure*}[t]
	\centering
	\includegraphics[scale=0.17]{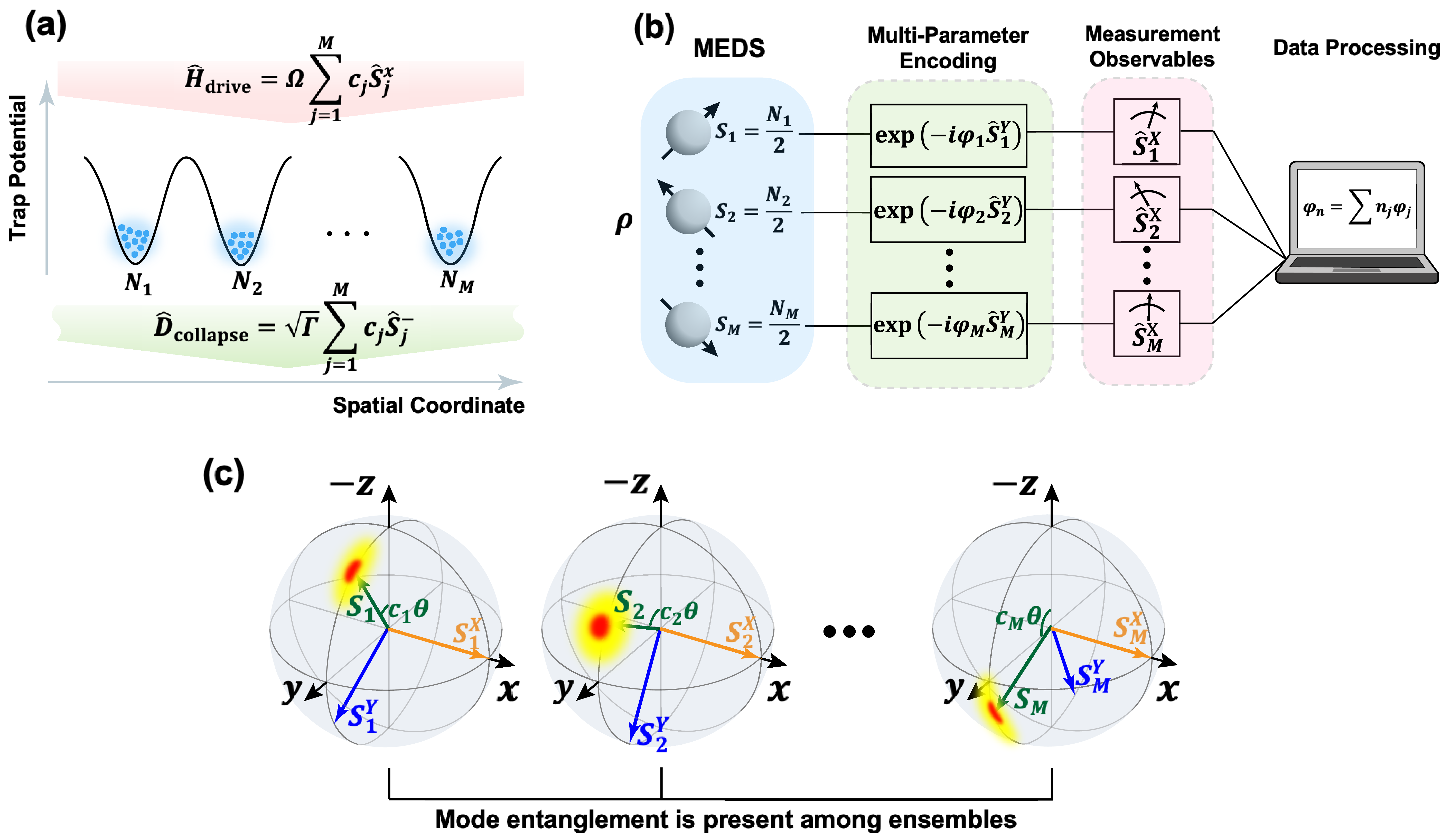}
	\caption{(a)~Schematic of a driven multi-ensemble superradiant system. A total of \(N\) identical particles are distributed into \(M\) spatially separated ensembles, which undergo collectively driven-dissipative dynamics. (b)~Spin-based multimode quantum interferometry. The unknown parameters \(\varphi_1, \ldots, \varphi_M\) are encoded onto the probe state MEDS via the unitary operator \(\hat{U}(\boldsymbol{\varphi}) = \exp(-\upi \boldsymbol{\varphi}^\top \hat{\mathbf{S}}^Y)\). The parameters are probed by measuring the observables \(\hat{S}_1^X, \ldots, \hat{S}_M^X\) ,and subsequently classically post-processed to infer a global parameter of interest. (c)~\(M\) Bloch spheres with mode entanglement. For the \(j\)th Bloch sphere (\(j=1,2,\ldots,M\)), the transverse spin components \(S_j^X\) and \(S_j^Y\) lie in the plane perpendicular to the Bloch vector \(\mathbf{S}_j\), with \(S_j^X\) chosen to be parallel to the global \(x\) axis. The Bloch vector \(\mathbf{S}_j\) forms an angle \(c_j\theta\) with respect to the \(-z\) axis, where \(\theta\) is a common parameter shared by all Bloch spheres. The squeezed modes within each ensemble, together with the metrologically useful inter-ensemble entanglement, enhance the measurement precision.}
	\label{Fig_1}
\end{figure*}

In this work, we introduce a driven multi-ensemble superradiant system as a natural extension of Dicke superradiance, designed to overcome the intrinsic limitations imposed by full permutation symmetry in single-ensemble settings. In contrast to the fully collective Dicke superradiance model, whose permutation symmetry constrains the complexity of both the transient dynamics and steady states~\cite{Anjun2025}, the multi-ensemble configuration explicitly breaks permutation symmetry between spatially separated ensembles. This symmetry breaking gives rise to multiple multi-ensemble dark states (MEDSs) whose structure and existence depend on the initial state of the system. Remarkably, these MEDSs exhibit a curvature-dependent enhancement of inter-ensemble entanglement~\cite{Shen2025,Sundar2024}, providing intrinsically generated quantum correlations that are well suited for metrological applications. Exploiting spatially separated MEDSs as probe states, we demonstrate enhanced precision in distributed quantum sensing, surpassing the SQL by fully harnessing the collective entanglement among the ensembles. The resulting quantum gain can be made optimal for arbitrary global phases by adjusting the system parameters, highlighting the flexibility of the multi-ensemble platform. We further note that the proposed multi-ensemble superradiant system can be realized in a driven–dissipative standing-wave cavity, where different ensembles experience distinct effective dissipation rates into the cavity mode due to variations in light–matter coupling strengths, thereby realizing the required symmetry breaking in a natural and experimentally accessible manner~\cite{Shen2025}.

The remainder of this paper is organized as follows. In Sec.~\ref{Sec.2}, we introduce both driven and undriven multi-ensemble superradiance. We analyze the multi-ensemble superradiant dynamics from the perspective of the superradiance potential and, within the Holstein–Primakoff approximation, derive analytical expressions for the covariance matrices of the MEDSs, together with a detailed characterization of their properties. In Sec.~\ref{Sec.3}, we introduce the multiparameter spin-squeezing coefficient, which can be viewed as a variational framework for optimizing quantum-enhanced sensitivity. This formulation allows us to identify the choice of system parameters under which it functions as a multimode quantum interferometer, achieving optimal estimation precision for arbitrary global parameters. We further account for finite-size effects and clarify the practical relevance of the proposed scheme beyond the thermodynamic limit. Finally, Sec.~\ref{Sec.4} summarizes the main results.

\section{Driven and Undriven Multi-ensemble Superradiance}\label{Sec.2}
We consider a driven, collectively dissipative two-level system consisting of a total of \(N\) identical particles partitioned into \(M\) spatially separated ensembles. As illustrated in Fig.~\ref{Fig_1}(a), the \(j\)th ensemble (\(j = 1, 2, \ldots, M\)) contains \(N_j = N \eta_j\) particles, where the population fractions satisfy \(\sum_{j=1}^{M} \eta_j = 1\). Each ensemble couples to the coherent drive and collective dissipation with a relative strength characterized by a normalized coefficient \(c_j\), subject to \(\sum_{j=1}^{M} c_j^2 = 1\). The driven-dissipative dynamics are governed by the master equation
\begin{equation}\label{drho_dt}
	\frac{\upd\hat{\rho}}{\upd t} = -\upi [\Omega \hat{\mathcal{S}}^x,\hat{\rho}] + \Gamma \mathcal{D}[\hat{\mathcal{S}}^{-}]\hat{\rho},
\end{equation}
where \(\Omega\) is the driving Rabi frequency and \(\Gamma\) is the collective decay rate. The dynamics are defined via the composite spin operators \(\hat{\mathcal{S}}^{\alpha} = \sum_{j=1}^{M} c_j \hat{S}_j^{\alpha}\) (\(\alpha = x,y,z,\pm\)) and the standard Lindblad superoperator \(\mathcal{D}[\hat{L}]\hat{\rho} = \hat{L} \hat{\rho} \hat{L}^\dagger - \frac{1}{2}\{ \hat{L}^\dagger\hat{L}, \hat{\rho} \}\). Here, the collective spin operators for the \(j\)th ensemble are \(\hat{S}_j^{\pm}=\sum_{i=1}^{N_j}{\hat{\sigma}_{j,i}^{\pm}}\), built from single-particle operators that satisfy \([\hat{\sigma}_{j,i}^{+},\hat{\sigma}_{j',i'}^{-}] = \hat{\sigma}_{j,i}^{z}\delta_{j,j'}\delta_{i,i'}\) and \([\hat{\sigma}_{j,i}^{z},\hat{\sigma}_{j',i'}^{\pm}] = \pm 2\hat{\sigma}_{j,i}^{\pm}\delta_{j,j'}\delta_{i,i'}\). In the absence of driving (\(\Omega = 0\)), the system reduces to undriven multi-ensemble superradiance. Importantly, unless the coupling coefficients \(c_j\) are identical, the composite operators \(\hat{\mathcal{S}}^{\alpha}\) do not satisfy the standard angular-momentum commutation relations. This departure from an underlying \(\mathrm{SU}(2)\) symmetry is central to the emergence of MEDSs without driving.

\subsection{Superradiance potential}
In the large-\(N\) limit, quantum fluctuations become negligible compared with expectation values, allowing the spin dynamics of each ensemble to be effectively described by a time-dependent Bloch equation~\cite{Sundar2024,Orioli2022,Shen2025},
\begin{equation}\label{dSj_dt}
	\frac{\upd}{\upd t}\mathbf{S}_j(t) = c_j \left[\Gamma \boldsymbol{\mathcal{S}}_{\perp}(t) + \mathbf{\Omega}\right] \times \mathbf{S}_j(t),
\end{equation}
where 
\(\mathbf{S}_j(t) = (\langle \hat{S}_j^x(t) \rangle, \langle \hat{S}_j^y(t) \rangle, \langle \hat{S}_j^z(t) \rangle)^\top\) 
denotes the Bloch vector of the \(j\)th ensemble as illustrated in Fig.~\ref{Fig_1}(c). 
The effective torque 
\(\Gamma \boldsymbol{\mathcal{S}}_{\perp}(t) + \mathbf{\Omega}\), 
with 
\(\boldsymbol{\mathcal{S}}_{\perp}(t)=(-\langle \hat{\mathcal{S}}^y(t) \rangle,\langle \hat{\mathcal{S}}^x(t) \rangle, 0)^\top\) 
and 
\(\mathbf{\Omega}=(\Omega,0,0)^\top\), 
governs the evolution of the Bloch vectors. For simplicity, we assume that, in the regime considered here, all Bloch vectors evolve within the \(y\)–\(z\) plane, such that the effective torque lies along the \(x\) axis. Equation~\eqref{dSj_dt} then implies that the angular velocity of each Bloch vector is proportional to its normalized coupling coefficient \(c_j\). Starting from the ground state, the positions of the Bloch vectors in the \(y\)–\(z\) plane can therefore be parameterized by a single collective variable \(\theta(t)\), with the angular displacement of the \(j\)th Bloch vector given by \(c_j \theta(t)\). 

To intuitively characterize the dynamics of \(\theta(t)\), and more importantly to highlight its intimate connection to the system’s correlation properties, we employ the superradiance potential \(V(\theta)\), defined through 
\(\upd \theta(t)/\upd t = -N\Gamma\, \upd V(\theta)/\upd \theta\), 
such that \(\theta(t)\) becomes stationary at the local minima of \(V(\theta)\)~\cite{Sundar2024,Orioli2022}. For the undriven multi-ensemble system, the potential takes the form (Appendix~\ref{Appendix A})
\begin{equation}
	V(\theta) = -\frac{1}{2}\sum_{j=1}^{M}{\eta_j \cos(c_j \theta)} + 2,
\end{equation}
where the constant term is irrelevant. A value of \(\theta\) corresponds to a minimum when the extremum and stability conditions, 
\(\upd V(\theta)/\upd\theta = 0\) and \(\upd^2V(\theta)/\upd\theta^2 > 0\), 
are satisfied. For Dicke superradiance, the corresponding superradiance potential reads 
\(V(\theta) = \frac{1}{2}(1-\cos{\theta})\). 
The photon emission rate can be identified with the negative rate of change of the excitation number, \(\gamma_{\mathrm{photon}} = -\upd\langle \hat{S}^z \rangle / \upd t\). 
Within the superradiance-potential framework, the radiation rate is proportional to the first derivative of \(V(\theta)\),
\begin{equation}
	\gamma_{\mathrm{photon}} = \frac{N^2\Gamma}{2}\sin{\theta}\frac{\upd V}{\upd\theta} = \frac{N^2\Gamma}{4}\sin^2{\theta},
\end{equation}
recovering the well-known \(N^2\) scaling of Dicke superradiance, with the maximal rate
\(\gamma_{\mathrm{photon}} = N^2\Gamma/4\) attained at \(\theta = \pi/2\). 
In the multi-ensemble superradiant system, imbalanced radiation amplitudes among different ensembles lead to destructive interference, driving the system into a MEDS. This state is characterized by a finite excitation number accompanied by a vanishing radiation rate, a feature that sharply distinguishes multi-ensemble superradiance from the Dicke case. These considerations establish a direct connection between the first derivative of the superradiance potential and the radiation rate of the system.

\begin{figure}[t]
	\centering
	\includegraphics[scale=0.38]{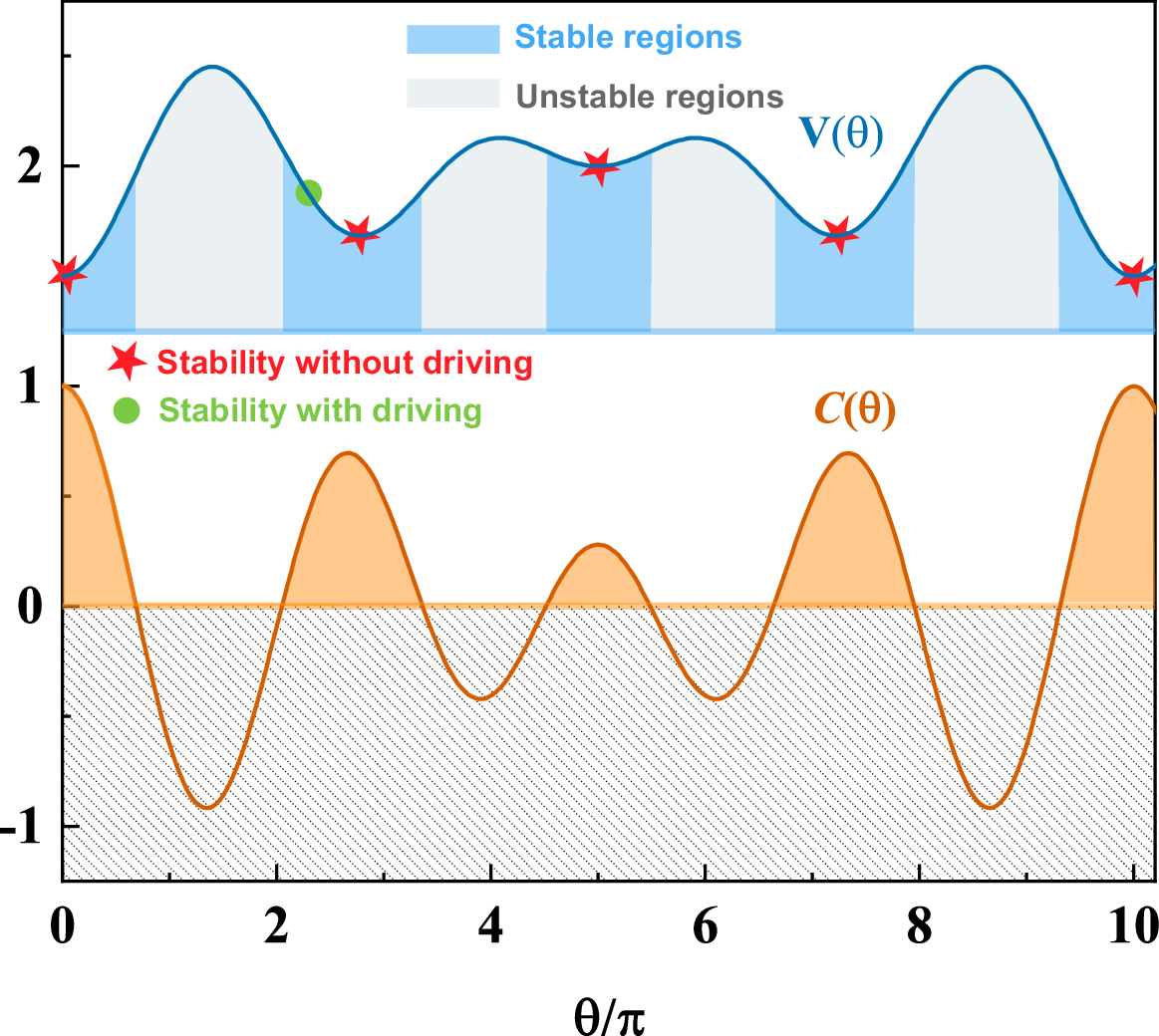}
	\caption{Superradiance potential \(V(\theta)\) and curvature \(\mathcal{C}(\theta)\) for the two-ensemble superradiant system, with \(\eta_1=\eta_2 = 1/2\), \(c_1 = 3/5\), and \(c_2 = 4/5\). In the undriven multi-ensemble superradiant system, stability is achieved only at discrete values of \(\theta\). However, with the introduction of a driving field, a steady state can be reached via the drive–dissipation balance within the stable region.}
	\label{Fig_2}
\end{figure}

Next, we introduce the curvature of the superradiance potential, which is proportional to the second derivative \(\upd^2V(\theta)/\upd\theta^2\). This quantity is closely related to the system’s correlation properties, in particular to quantum entanglement that is beneficial for quantum metrology. The curvature is defined as
\begin{equation}\label{C_theta}
	\mathcal{C}(\theta) = \frac{\mathcal{C}_{\theta}}{\mathcal{C}_0},
\end{equation}
where 
\begin{equation}
	\mathcal{C}_{\theta} = \sum_{j=1}^{M}{\eta_j c_j^2 \cos(c_j\theta)},
\end{equation}
and the normalization factor 
\(\mathcal{C}_0 = \sum_{j=1}^{M}\eta_j c_j^2\) is chosen such that \(\lvert \mathcal{C}(\theta) \rvert \le 1\). The normalized curvature \(\mathcal{C}(\theta)\) evaluated at the minima of the potential quantifies the local steepness of the potential well. As shown in Fig.~\ref{Fig_2}, in the absence of driving the system becomes stationary only at discrete values of \(\theta\), with the corresponding curvature determined by the system parameters \(\{\eta_j, c_j\}\). This discreteness restricts the accessible steady-state curvatures. 

In contrast, under external driving the system can reach a steady state at any \(\theta\) within the stable region through a balance between drive and dissipation (Appendix~\ref{Appendix A}), 
\begin{equation}
	\Omega(\theta) = \frac{\Gamma}{2}\sum_{j=1}^{M}{N_j c_j \sin(c_j \theta)}.
\end{equation}
From a macroscopic geometric perspective, the introduction of this external coherent drive effectively superimposes a linear potential upon the system. This addition tilts the original undriven superradiance potential, dynamically reshaping it into a ``tilted washboard'' configuration. In this view, the local minima, which dictate the accessible steady states, are continuously shifted by the external drive. The driving field therefore provides an additional degree of freedom for controlling the steady-state properties of the multi-ensemble superradiant system.

To prepare the MEDS corresponding to a desired curvature, we initialize the system in a \(\theta\)-dependent product state,
\begin{equation}\label{psi_theta}
	\lvert \psi_\theta \rangle = \bigotimes_{j=1}^M\left[ \cos\!\left(\frac{c_j \theta}{2}\right) \lvert g \rangle -\upi\sin\!\left(\frac{c_j \theta}{2}\right)\lvert e \rangle \right]^{\otimes N_j},
\end{equation}
where the value of \(\theta\) is determined by Eq.~\eqref{C_theta}. This initial state can be generated by applying the unitary operator 
\(\hat{U} = \exp{(-\upi\theta\hat{\mathcal{S}}^x)}\) 
to the ground state \(\lvert g \rangle^{\otimes N}\)~\cite{Shen2025}. By preparing the system in the separable state \(\lvert \psi_\theta \rangle\) and simultaneously tuning the external drive to precisely balance the collective dissipation, we establish a rigorous driven-dissipative equilibrium right from the initial time. Since the system is macroscopically initialized exactly at the potential minimum dictated by \(\theta\), the Bloch vectors (first moments) of all ensembles remain approximately frozen near their initial orientations. Instead, the continuous interplay between driving and dissipation exclusively drives the quantum fluctuations (second moments) to dynamically adapt to the local geometry of this minimum. This specific evolution of the second moments signals the transition from an uncorrelated product state to a highly entangled MEDS. As we will rigorously demonstrate in the subsequent sections, the ultimate quantum correlations of this generated dark state are intrinsically governed by the local curvature.

\subsection{Dissipative dynamics in the bosonic representation}
Analyzing spin systems directly is challenging due to the nontrivial commutation relations of spin operators. The Holstein–Primakoff transformation provides a powerful approximation by mapping spin operators onto bosonic modes, which becomes accurate in the large-spin or low-excitation limit~\cite{Shen2025,Kurucz2010,Holstein1940,Ma2011}. When the system reaches a driven–dissipative equilibrium at the initial time, the Bloch vectors remain frozen near their initial positions. Since the Bloch vectors lie within the laboratory \(y\)-\(z\) plane, we introduce a local rotated coordinate frame (\(X\),\(Y\),\(Z\)) for each ensemble individually. Specifically, the rotated \(X\)-axis is defined to be parallel to the laboratory \(x\)-axis for all ensembles. However, the \(Y\) and \(Z\) axes are rotated about the \(x\)-axis such that the negative local \(Z\)-axis aligns with the initial mean spin vector of the specific ensemble. Consequently, the directions of the \(Y\) and \(Z\) axes generally differ from ensemble to ensemble, depending on their respective initial-state orientations. Accordingly, the system's initial state corresponds to the ground state of the spin operators \(\hat{S}_j^Z\) (pointing along −Z) for each subsystem. To facilitate the subsequent analysis in the bosonic representation, we note that the master equation~\eqref{drho_dt} can be exactly recast into a purely dissipative form, \(\upd\hat{\rho}/\upd t=\Gamma \mathcal{D}[\hat{D}^-]\hat{\rho}\), by introducing an effective collective lowering operator \(\hat{D}^{-} \equiv \upi\Omega/\Gamma + \hat{\mathcal{S}}^{-}\). In the rotated basis, the spin operators can be mapped onto bosonic operators via the Holstein–Primakoff approximation as (Appendix~\ref{Appendix B})
\begin{equation}\label{eq:H-P_app}
	\left\{
	\begin{aligned}
		&\hat{S}_j^X + \upi \hat{S}_j^Y \sim \sqrt{N_j}\hat{a}_j^\dagger,\\
		&\hat{S}_j^X - \upi \hat{S}_j^Y \sim \sqrt{N_j}\hat{a}_j,\\
		&\hat{S}_j^Z \sim - \frac{N_j}{2},
	\end{aligned}
	\right.
\end{equation}
where \(\hat{S}_j^X\) and \(\hat{S}_j^Y\) denote the collective spin components along the previously defined local \(X\) and \(Y\) axes, and the bosonic operators \(\hat{a}_j\) and \(\hat{a}_j^\dagger\) satisfy the canonical commutation relation \([\hat{a}_j, \hat{a}_k^\dagger] = \delta_{jk}\). 

In the bosonic representation, the initial state corresponds to the ground state of the \(M\) bosonic modes, i.e., the multimode vacuum \(\lvert \psi_\theta \rangle = \lvert 0 \rangle^{\otimes M}\). The effective collective lowering operator can then be written as a multimode Bogoliubov operator, 
\begin{equation}
	\hat{D}^{-} = \sqrt{\mathcal{C}_{\theta} N}\hat{C}_1,
\end{equation}
where
\begin{equation}
	\hat{C}_1 = \sum_{j=1}^{M}{\sqrt{\frac{\eta_j}{\mathcal{C}_{\theta}}}c_j\left[\cos^2\!\left(\frac{c_j \theta}{2}\right)\hat{a}_j + \sin^2\!\left(\frac{c_j \theta}{2}\right)\hat{a}_j^\dagger\right]}.
\end{equation} 
In this representation, the master equation~\eqref{drho_dt} reduces to
\begin{equation}
	\frac{\upd\hat{\rho}}{\upd t} = \mathcal{C}_{\theta} N \Gamma \left(\hat{C}_1\hat{\rho}\,\hat{C}_1^\dagger-\frac{1}{2}\hat{\rho}\,\hat{C}_1^\dagger\hat{C}_1- \frac{1}{2}\hat{C}_1^\dagger\hat{C}_1\hat{\rho}\right).
\end{equation}
This formulation reveals that the collective mode associated with \(\hat{C}_1\) is dissipatively cooled to its vacuum state, thereby generating entanglement among the ensembles, while the remaining collective modes remain dynamically frozen and retain memory of their initial conditions.

\subsection{Dark-state covariance matrix and its key properties}
The covariance matrix characterizes quantum fluctuations and correlations among observables in a multimode or many-body system, and serves as a central tool for detecting entanglement and optimizing measurement precision in quantum metrology. To analyze the dark-state covariance matrices, we introduce the quadrature operators collected in the vector 
\(\hat{\mathbf{Q}}^a = (\hat{X}_1^a, \ldots, \hat{X}_M^a, \hat{Y}_1^a, \ldots, \hat{Y}_M^a)^\top\), 
where 
\(\hat{X}_j^a = (\hat{a}_j + \hat{a}_j^\dagger)/\sqrt{2}\) and \(\hat{Y}_j^a = \upi(\hat{a}_j^\dagger -\hat{a}_j)/\sqrt{2}\) 
satisfy the canonical commutation relations 
\([\hat{X}_j^a, \hat{Y}_k^a] = \upi\delta_{jk}\). 
The covariance matrix \(\bm{\Gamma}_Q^a\) is defined by 
\(\left(\bm{\Gamma}_Q^a\right)_{jk} = \langle \hat{Q}_j^a \hat{Q}_k^a + \hat{Q}_k^a \hat{Q}_j^a \rangle\). Here we adopt a convention without the factor of \(1/2\), which is convenient for the present analysis. For the MEDS associated with a given \(\theta\), the dark-state covariance matrix admits the analytic form (Appendix~\ref{Appendix C})
\begin{equation}
	\bm{\Gamma}_Q^a(\infty) = \left(\mathbbm{1}_M - \widetilde{\bm{\Gamma}}_X^a \right)\oplus\left(\mathbbm{1}_M - \widetilde{\bm{\Gamma}}_Y^a \right),
\end{equation}
where \(\mathbbm{1}_M\) denotes the \(M\times M\) identity matrix. The correction matrices are given by
\begin{equation}
	\widetilde{\bm{\Gamma}}_X^a = \mathbf{A}\mathbf{B}^\top + \mathbf{B}\mathbf{A}^\top - \left(1 + \lVert \mathbf{A} \rVert^2 \right)\mathbf{B}\mathbf{B}^\top,
\end{equation} 
and
\begin{equation}
	\widetilde{\bm{\Gamma}}_Y^a= \mathbf{A}\mathbf{B}^\top + \mathbf{B}\mathbf{A}^\top - \left(1 + \lVert \mathbf{B} \rVert^2 \right)\mathbf{A}\mathbf{A}^\top,
\end{equation} 
with 
\(\mathbf{A}=\sqrt{\frac{1}{\mathcal{C}_{\theta}}}\left(c_1\sqrt{\eta_1}, \ldots, c_M\sqrt{\eta_M}\right)^\top\) and \(\mathbf{B}=\sqrt{\frac{1}{\mathcal{C}_{\theta}}}\left(c_1\sqrt{\eta_1}\cos(c_1\theta),\ldots,c_M\sqrt{\eta_M}\cos(c_M\theta)\right)^\top\). The inner product satisfies \(\mathbf{A}^\top\mathbf{B}\equiv 1\), while
\(\|\mathbf{A}\|=\mathcal{C}^{-1/2}\).
The angle \(\phi\) between the two vectors is defined via
\(\sec\phi=\|\mathbf{A}\|\|\mathbf{B}\|\),
with the inequalities
\(\|\mathbf{A}\|^2 \ge \sec\phi \ge \|\mathbf{B}\|^2\).

The minimum eigenvalue of the dark-state covariance matrix is given by (Appendix~\ref{Appendix D})
\begin{equation}\label{lambda_min}
	\lambda_{\mathrm{min}} = \frac{1}{1+\cos(2\phi)}\left[ 1 + \mathcal{C} - \sqrt{1+\mathcal{C}^2 - 2 \mathcal{C} \cos(2\phi)} \right],
\end{equation}
which is real, positive, and does not exceed unity. As the curvature \(\mathcal{C}\) decreases, \(\lambda_\mathrm{min}\) decreases correspondingly and scales as \(\lambda_\mathrm{min} \sim \mathcal{C}\) in the limit \(\mathcal{C} \ll 1\), vanishing as \(\mathcal{C} \to 0\). Thus, MEDSs with lower curvature give rise to dark-state covariance matrices with smaller minimum eigenvalues. The eigenvector corresponding to the minimum eigenvalue is
\begin{equation}\label{v_min}
	\mathbf{v}_\mathrm{min} = \cos \! \left(\frac{\beta}{2}\right)\mathbf{e}_1 - \sin \! \left(\frac{\beta}{2}\right) \mathbf{e}_2,
\end{equation}
where \(\beta \in [ 0, \pi/2]\) is defined via \(\tan\beta = \sin (2\phi)/\left[\mathcal{C}^{-1} - \cos (2\phi)\right]\). The two orthogonal unit vectors \(\mathbf{e}_1\) and \(\mathbf{e}_2\) span the two-dimensional subspace defined by \(\mathbf{A}\) and \(\mathbf{B}\), with \(\mathbf{e}_1 \parallel \mathbf{A}\) and \(\mathbf{e}_2\) orthogonal to \(\mathbf{A}\) and having a positive overlap with \(\mathbf{B}\). In the limit \(\mathcal{C} \ll 1\), the asymptotic behavior \(\cos(\beta/2)\sim 1-\mathcal{C}^2\sin^2(2\phi)/8\) and \(\sin(\beta/2)\sim\mathcal{C}\sin(2\phi)/2\) shows that the eigenvector becomes increasingly aligned with \(\mathbf{A}\) as the curvature decreases.

The difference between the \(X\)-quadrature block and the inverse of the \(Y\)-quadrature block of the dark-state covariance matrix is given by
\begin{equation}\label{difference}
	\left(\mathbbm{1}_M - \widetilde{\bm{\Gamma}}_X^a\right) - \left(\mathbbm{1}_M - \widetilde{\bm{\Gamma}}_Y^a\right)^{-1} = \tan^2\!\phi \, \mathbf{e}_2\mathbf{e}_2^\top.
\end{equation} 
In particular, this difference vanishes when \(\mathbf{A}\) and \(\mathbf{B}\) become parallel. Physically, this condition indicates that the steady state approaches a pure minimum-uncertainty Gaussian state. This property is highly advantageous for quantum metrology: first, the high purity ensures the absence of classical statistical noise, thereby maximizing the quantum metrological gain quantified by the quantum Fisher information matrix (QFIM); second, the Gaussian nature of the state implies that the ultimate precision limit, the quantum Cram\'er-Rao bound, can generally be saturated via specific, practically accessible measurement schemes. We provide a detailed derivation of the QFIM and discuss its connection to the covariance matrix in Appendix~\ref{Appendix F}.

\section{Spin-based multimode quantum interferometry}\label{Sec.3}
Before detailing the interferometry protocol, it is necessary to explicitly establish the connection between the steady-state solutions obtained in Sec.~\ref{Sec.2} C and the quantum metrological bounds discussed below. Under the Holstein-Primakoff approximation, the driven-dissipative dynamics prepare the distributed ensembles in a collective Gaussian state. A fundamental property of Gaussian states is that all of their quantum correlations and fluctuation statistics are completely and exclusively encapsulated by their first moments and the covariance matrix. Consequently, the optimal spin squeezing and the ultimate precision limits of our distributed quantum sensor are entirely determined by this bosonic covariance matrix, making it the central mathematical object for the following metrological analysis.

In distributed quantum sensing, the objective is typically not to estimate each local parameter individually, but rather to infer a global quantity encoded across spatially distributed subsystems. Such a global parameter can often be expressed as a linear combination of the local parameters,
\begin{equation}
	\varphi_{\mathbf n}
	:= \mathbf n^\top \boldsymbol{\varphi}
	= \sum_{j=1}^{M} n_j \varphi_j,
\end{equation}
where \(\boldsymbol{\varphi}=(\varphi_1,\varphi_2,\ldots,\varphi_M)^\top\) denotes the vector of local parameters, and \(\mathbf n\in\mathbb R^M\) specifies the estimation direction, normalized such that \(\sum_{j=1}^{M}|n_j|=1\).
This formulation naturally captures a wide range of distributed sensing tasks, including the estimation of phase differences in spatially separated interferometers~\cite{Aasi2013,Knott2016}, the determination of global averages or sums of multiple parameters~\cite{Komar2014}, and the sensing of spatial gradients~\cite{Zhang2014,Ng2014}. 

In the driven multi-ensemble superradiant system, as illustrated in Fig.~\ref{Fig_1}(b), we employ the MEDS as the probe state for spin-based multimode quantum interferometry. The unknown phases \(\boldsymbol{\varphi}\) are locally encoded onto the probe state via the unitary evolution 
\(\hat{U}(\boldsymbol{\varphi}) = \exp{(-\upi \boldsymbol{\varphi}^\top \hat{\mathbf{S}}^Y)} = \exp{(-\upi \sum_{j=1}^{M} \varphi_j \hat{S}_j^Y)}\). The phases are subsequently inferred by measuring the spin components
\(\hat{\mathbf{S}}^X = (\hat{S}_1^X,\ldots,\hat{S}_M^X)^\top\) of the respective ensembles, with the measurement outcomes classically post-processed to estimate a global parameter of interest. It is important to note a practical geometric feature of this encoding scheme. According to our local frame definition, while the optimally squeezed local \(X\)-axes are all parallel to the laboratory \(x\)-axis, the anti-squeezed local \(Y\)-axes lie at different angles within the laboratory \(y\)-\(z\) plane. Consequently, directly encoding parameters via the generators \(\hat{S}_j^Y\) would require external driving fields to be fine-tuned along different directions for each ensemble. However, this practical limitation can be elegantly circumvented by applying suitable local unitaries prior to the sensing phase. Mathematically, the local operator \(\hat{S}_j^Y\) is related to the laboratory-frame operator \(\hat{S}_j^y\) via a local rotation \(\hat{R}_j\) about the laboratory \(x\)-axis, such that \(\hat{S}_j^Y = \hat{R}_j \hat{S}_j^y \hat{R}_j^\dagger\). The desired encoding can therefore be equivalently implemented by first applying local unitaries \(\hat{R}_j^\dagger\) to align all local \(Y\)-axes with the laboratory \(y\)-axis, followed by a spatially distributed multi-parameter sensing field pointing purely along the laboratory \(y\)-axis, and finally applying \(\hat{R}_j\) to restore the ensembles to their original local frames. Because the measurement directions (along the local \(X\)-axes) are naturally aligned with the laboratory \(x\)-axis, the subsequent optimal readout can be performed globally without further local manipulations. 

For each ensemble, the operators \(\hat{S}_j^X\) and \(\hat{S}_j^Y\) lie in the plane orthogonal to the mean spin direction. At the shot-noise limit, the QFIM associated with the generators \(\hat{\mathbf{S}}^Y\) takes the diagonal form \(\bm{F}_\mathrm{SN}[\hat{\mathbf{S}}^Y] = \mathrm{diag}(N_1,N_2,\ldots,N_M)\)~\cite{Gessner2018}. 
For a given estimation direction \(\mathbf{n}\), the corresponding phase sensitivity at the shot-noise limit is 
\((\Delta \varphi_\mathbf{n})^2_{\mathrm{SN}} = \mathbf{n}^\top \bm{F}_\mathrm{SN}^{-1} \mathbf{n}\). 
This sensitivity can be optimized by appropriately distributing the particles among the ensembles while keeping the total particle number fixed. Using the method of Lagrange multipliers, the optimal shot-noise–limited sensitivity,
\((\Delta \varphi_{\mathbf n})^2_{\mathrm{SN,opt}}=N^{-1}\),
which coincides with the SQL, is obtained for population fractions \(\eta_j=|n_j|\).

In the multiparameter scenario, we introduce an analog of the spin-squeezing coefficient~\cite{Wineland1992} from single-parameter estimation, defined as (Appendix~\ref{Appendix E})
\begin{equation}
	\xi_\mathbf{n}^2[\hat{\mathbf{S}}^Y,\hat{\mathbf{S}}^X] := \frac{\left(\Delta \varphi_\mathbf{n}\right)^2}{\left(\Delta \varphi_\mathbf{n}\right)^2_\mathrm{SN}} = \mathcal{R}\left( \bm{\Xi}^2[\hat{\mathbf{S}}^Y,\hat{\mathbf{S}}^X], \mathbf{m} \right),
\end{equation} 
where \(\mathcal{R}(\bm{\Xi}^2[\hat{\mathbf{S}}^Y,\hat{\mathbf{S}}^X], \mathbf{m}) = (\mathbf{m}^\top \bm{\Xi}^2[\hat{\mathbf{S}}^Y,\hat{\mathbf{S}}^X] \mathbf{m}) / (\mathbf{m}^\top \mathbf{m})\)
denotes the \emph{Rayleigh quotient} of the spin-squeezing matrix \(\bm{\Xi}^2[\hat{\mathbf{S}}^Y,\hat{\mathbf{S}}^X]\), 
evaluated along the transformed vector 
\(\mathbf{m} = \bm{F}^{-1/2}_{\mathrm{SN}}[\hat{\mathbf{S}}^Y] \mathbf{n}\). 
The resulting multiparameter spin-squeezing coefficient (MSSC) is bounded by the minimum and maximum eigenvalues of the spin-squeezing matrix and attains these extrema when \(\mathbf{m}\) aligns with the corresponding eigenvectors~\cite{Horn2012}. Accordingly, the MSSC probes the spectrum of the spin-squeezing matrix along the direction specified by \(\mathbf m\), and can be interpreted as a variational principle: by tuning the system parameters such that the eigenvector associated with the minimum eigenvalue of the spin-squeezing matrix aligns with the fixed direction \(\mathbf m\), the MSSC is minimized, thereby yielding optimal quantum gain for the estimation of arbitrary linear combinations of parameters.

The spin-squeezing matrix, introduced by Gessner \emph{et al.}~\cite{Gessner2020}, is defined as
\begin{equation}\label{eq:Xi^2}
	\left( \bm{\Xi}^2[\hat{\mathbf{S}}^Y,\hat{\mathbf{S}}^X] \right)_{jk} = \frac{\sqrt{N_j N_k} \, \mathrm{Cov} \! \left(\hat{S}_j^X, \hat{S}_k^X\right)}{\langle \hat{S}_j^Z \rangle \langle \hat{S}_k^Z \rangle},
\end{equation} 
where \(\mathrm{Cov} (\hat{S}_j^X, \hat{S}_k^X) = \frac{1}{2}\langle \hat{S}_j^X\hat{S}_k^X + \hat{S}_k^X\hat{S}_j^X \rangle - \langle \hat{S}_j^X \rangle \langle \hat{S}_k^X \rangle\) denotes the covariance of the measurement observables. To rigorously evaluate this spin-squeezing matrix using our steady-state solutions, we utilize the Holstein-Primakoff transformation established in Eq.~\eqref{eq:H-P_app}. By directly substituting the macroscopic collective spin operators in Eq.~\eqref{eq:Xi^2} with their corresponding bosonic quadrature operators, we map the spin-squeezing matrix onto the covariance matrix of the collective bosonic modes. This straightforward algebraic substitution directly yields the expression based on the bosonic covariance matrix:
\begin{equation}
	\bm{\Xi}^2[\hat{\mathbf{S}}^Y,\hat{\mathbf{S}}^X] = \mathbbm{1}_M - \widetilde{\bm{\Gamma}}_X^a.
\end{equation} 
Its minimum eigenvalue \(\lambda_{\mathrm{min}}\), given in Eq.~\eqref{lambda_min}, is always smaller than unity, signaling the presence of multiparameter spin squeezing. In the following, we present a concrete protocol for achieving optimal quantum gain along an arbitrary estimation direction \(\mathbf n\) by appropriately tuning the population fractions \(\eta_j\) and the relative couplings \(c_j\) of the ensembles.

Consider uniform coupling, i.e., \(\lvert c_1 \rvert = \lvert c_2 \rvert = \cdots = \lvert c_M \rvert = 1/\sqrt{M}\), for which the vectors \(\mathbf{A}\) and \(\mathbf{B}\) are parallel (\(\phi=0\)). In this case, the minimum eigenvalue of the spin-squeezing matrix is
\begin{equation}\label{eq:lamdaC}
	\lambda_\mathrm{min} = \mathcal{C}(\theta),
\end{equation}
which directly coincides with the curvature of the superradiance potential confining the MEDS. The associated eigenvector is \(\mathbf{v}_{\mathrm{min}} = \left(\mathrm{sgn}(c_1)\sqrt{\eta_1}, \ldots, \mathrm{sgn}(c_M)\sqrt{\eta_M} \right)^\top\). The explicit analytical form of this highly symmetric squeezing matrix is detailed in Appendix~\ref{Appendix E}. Physically, the inverse relationship between the potential curvature and the metrological sensitivity, as implied by Eq.~\eqref{eq:lamdaC}, can be intuitively understood via the Heisenberg uncertainty principle. A smaller curvature implies a flatter superradiance potential, which weakly confines the system and leads to large quantum fluctuations in the collective angular coordinate \(\theta\). Through linear error propagation on the Bloch sphere, the rotation around the \(\hat{S}_j^X\) axis couples this large angular uncertainty directly into the orthogonal \(\hat{S}_j^Y\) quadrature, manifesting as massive anti-squeezing. Because the system reaches a minimum-uncertainty pure dark state in the uniform coupling regime, this large anti-squeezing in \(\hat{S}^Y\) strictly enforces a strong quantum squeezing in the orthogonal measurement observable \(\hat{S}^X\) (i.e., \(\Delta \hat{S}^X \Delta \hat{S}^Y \sim \text{const}\)). However, we emphasize that this simplistic geometric dependence breaks down in the non-uniform coupling case. In that regime, the disparate coupling strengths and the resulting unbalanced effective radiation rates prevent the system from cooling into a pure state. Instead, the system relaxes into a mixed state, and the quantum gain is compromised by the introduced classical mixture. Consequently, the generalized uncertainty relation is no longer saturated (\(\Delta \hat{S}^X \Delta \hat{S}^Y > \text{const}\)), and the metrological performance can no longer be governed by Eq.~\eqref{eq:lamdaC}, but must instead be rigorously evaluated using the comprehensive covariance matrix formalism presented in Eq.~\eqref{lambda_min}. Furthermore, as shown in Appendix~\ref{Appendix E}, the spin-squeezing coefficient of the \(j\)th ensemble is \(1-\eta_j(1-\mathcal{C})\), which generally exceeds the global curvature \(\mathcal{C}\). This explicitly signals that the ultimate overall measurement precision is not merely a sum of local effects, but relies fundamentally on the presence of metrologically useful inter-ensemble entanglement that coordinates the distributed nodes to further enhance the overall measurement precision.

To achieve the optimal MSSC, the transformed vector \(\mathbf{m}\) must be parallel to the eigenvector associated with the minimum eigenvalue of the spin-squeezing matrix. This requirement yields
\begin{equation}
	\left\{\eta_j, c_j \right\} = \left\{ \lvert n_j \rvert, \frac{\mathrm{sgn}(n_j)}{\sqrt{M}}\right\},
\end{equation}
for which the population fractions coincide with those that optimize the shot-noise–limited sensitivity. Consequently, for a fixed estimation direction \(\mathbf n\), the phase sensitivity
\begin{equation}
	(\Delta \varphi_{\mathbf n})^2=\frac{\mathcal{C}}{N}
\end{equation}
is optimal among all possible choices of particle-number distributions \(\{N_j\}\), since any other distribution effectively reduces the total contributing particle number below \(N\).

\begin{figure}[t]
	\centering
	\includegraphics[scale=0.37]{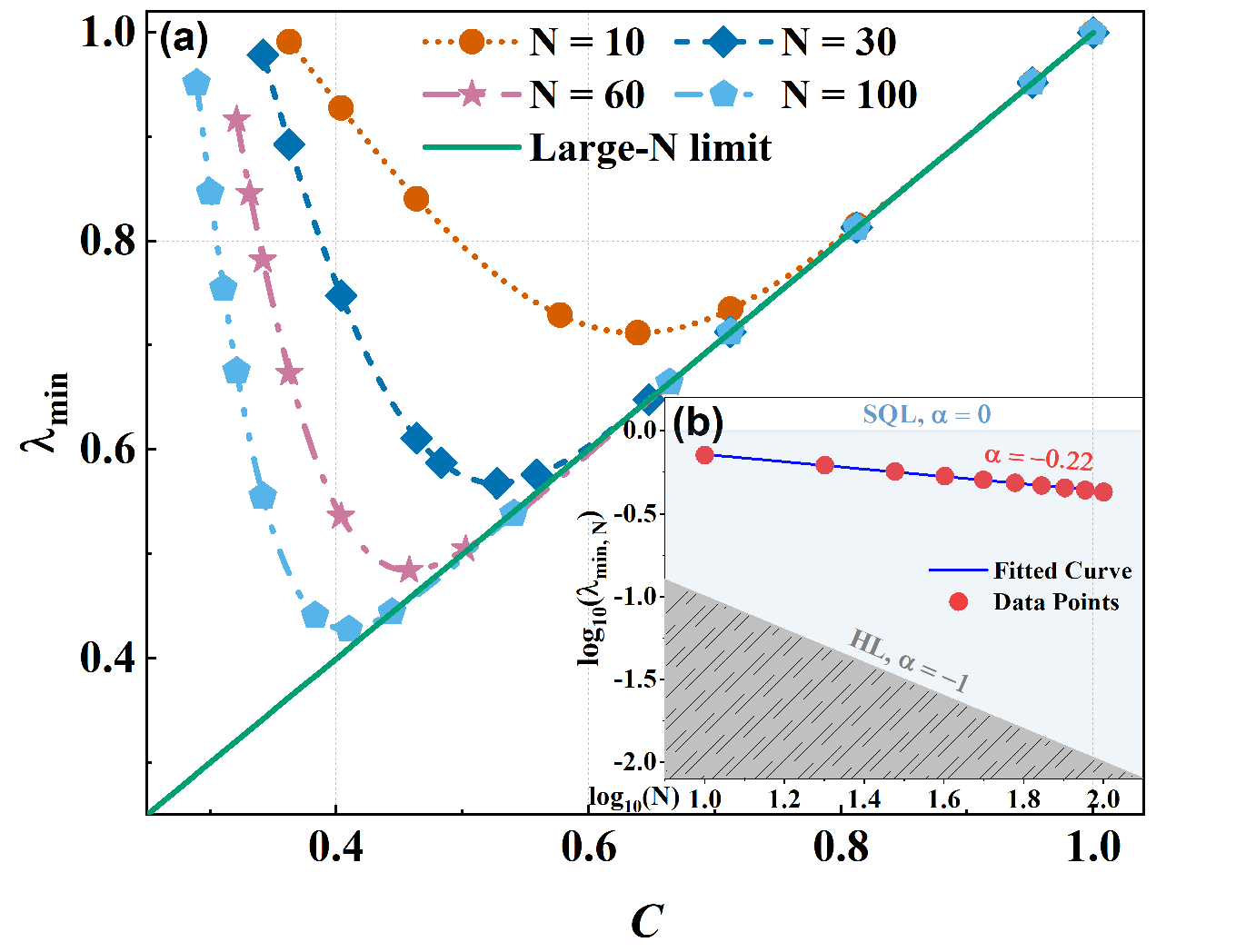}
	\caption{(a)~In the uniform coupling case, the minimum eigenvalue \(\lambda_{\mathrm{min}}\) of the spin-squeezing matrix as a function of curvature \(\mathcal{C}\) for different particle numbers. (b)~The logarithm of \(\lambda_{\mathrm{min},N}\) as a function of the logarithm of particle numbers \(N\). Here, \(\lambda_{\mathrm{min},N}\) denotes the smallest value of \(\lambda_{\mathrm{min}}\) for all curvatures at a given \(N\). The exponent \(\alpha\) is defined via \(\lambda_{\mathrm{min},N} \propto N^{\alpha}\). \(\alpha = 0\) and \(\alpha = -1\) correspond to the standard quantum limit (SQL) and the Heisenberg limit (HL), respectively. In these simulations, the potential curvature \(\mathcal{C}\) is tuned by varying the strength of the coherent drive. It is important to note that for a fixed total particle number \(N\), the plotted ultimate optimal quantum gain (derived from \(\lambda_{\min}\)) is completely invariant to the specific particle number ratio \(\eta_j\) among the ensembles. Therefore, these results are universally applicable to any arbitrary atomic distribution.}
	\label{Fig_3}
\end{figure}

In general, although the spin-squeezing matrix admits a simple analytical form, its construction relies solely on first- and second-order moments and therefore does not provide a complete characterization of a system’s quantum-metrological performance. In our protocol, the probe state (MEDS) is pure. Moreover, as we show below, the spin-squeezing matrix fully captures the relevant quantum-metrological features (Appendix~\ref{Appendix F}). We introduce a modified spin-squeezing matrix based on the upper bound of the QFIM for pure probe states, given by \(4\bm{\Gamma}[\hat{\mathbf{S}}^Y]\). In the present setting, this modified matrix takes the form 
\((\mathbbm{1}_M - \widetilde{\bm{\Gamma}}_Y^a)^{-1}\), 
which coincides with the spin-squeezing matrix 
\(\mathbbm{1}_M - \widetilde{\bm{\Gamma}}_X^a\) 
when \(\phi = 0\), as shown in Eq.~\eqref{difference}. This equivalence establishes that the choice of \(\{\hat{\mathbf{S}}^Y,\hat{\mathbf{S}}^X\}\) constitutes an optimal strategy for the pure probe states corresponding to uniform coupling.

In Fig.~\ref{Fig_3}(a), we consider a driven two-ensemble superradiant system with uniform coupling. To obtain the numerical values of the optimal quantum gain \(\lambda_{\min}\), we first numerically integrate the full Lindblad master equation in Eq.~\eqref{drho_dt} using the open-source Quantum Toolbox in Python (QuTiP)~\cite{Johansson2012} to extract the steady-state first- and second-order spin moments. These moments are subsequently substituted into Eq.~\eqref{eq:Xi^2} to construct the squeezing matrix, whose minimum eigenvalue is then numerically evaluated to yield \(\lambda_{\min}\). It is crucial to highlight the decoupled physical dependencies governing the distributed parameter estimation. Our results reveal that the ultimate optimal quantum gain, \(\lambda_{\min}\), relies only on the total particle number \(N\) and the superradiance potential curvature \(\mathcal{C}\) (tuned by the coherent drive), and is remarkably invariant to the specific atomic distribution \(\eta_j = N_j/N\). However, to actually achieve this universal optimal gain, the target estimation direction \(\mathbf{n}\) must perfectly match the resource allocation, satisfying \(\eta_j = |n_j|\). Consequently, the specific optimal linear combination being estimated is completely dictated by the chosen distribution ratio \(\eta_j\), and does not vary with either the curvature \(\mathcal{C}\) or the total particle number \(N\). Although our analytical results are derived in the large-\(N\) limit, numerical simulations show excellent agreement even for modest atom numbers, provided that the curvature is not too small. When finite-size effects are taken into account, we find that, as the curvature \(\mathcal{C}\) decreases, the minimum eigenvalue \(\lambda_{\mathrm{min}}\) gradually deviates from \(\mathcal{C}\). Nevertheless, for larger particle numbers this deviation sets in only at progressively smaller values of \(\mathcal{C}\). 
This behavior can be understood as follows: in finite-size systems, significant quantum fluctuations persist within the superradiance potential well at small curvatures, allowing the MEDS to escape. As the particle number increases, quantum fluctuations are further suppressed, and escape becomes possible only for increasingly shallower potential wells. In Fig.~\ref{Fig_3}(b), we plot \(\lambda_{\mathrm{min},N}\), defined as the minimum of \(\lambda_{\mathrm{min}}\) over all curvatures, for various particle numbers. The data reveal a finite-size scaling \(\lambda_{\mathrm{min},N} \propto N^{-0.22}\), reflecting the competition between quantum fluctuations and the superradiance potential.

Beyond the ideal uniform limit, our generalized multi-ensemble framework proves practically essential for evaluating realistic experimental platforms. For instance, in a standing-wave cavity QED system, spatially separated atomic ensembles experience position-dependent coupling strengths, typically \(g_j \propto \cos(k x_j)\). In our theory, this spatial dependence maps directly to the inhomogeneous coefficients \(c_j\). This explicit mapping allows us to rigorously capture how such inhomogeneities break the collective permutation symmetry and force the system into a mixed state contaminated by classical statistical uncertainty. While simplistic collective spin models completely fail to describe this classical mixedness, our generalized covariance matrix formalism successfully quantifies the surviving quantum gain, thereby defining the practical operational regime resilient to experimental imperfections.

\begin{figure}[t]
	\centering
	\includegraphics[scale=0.36]{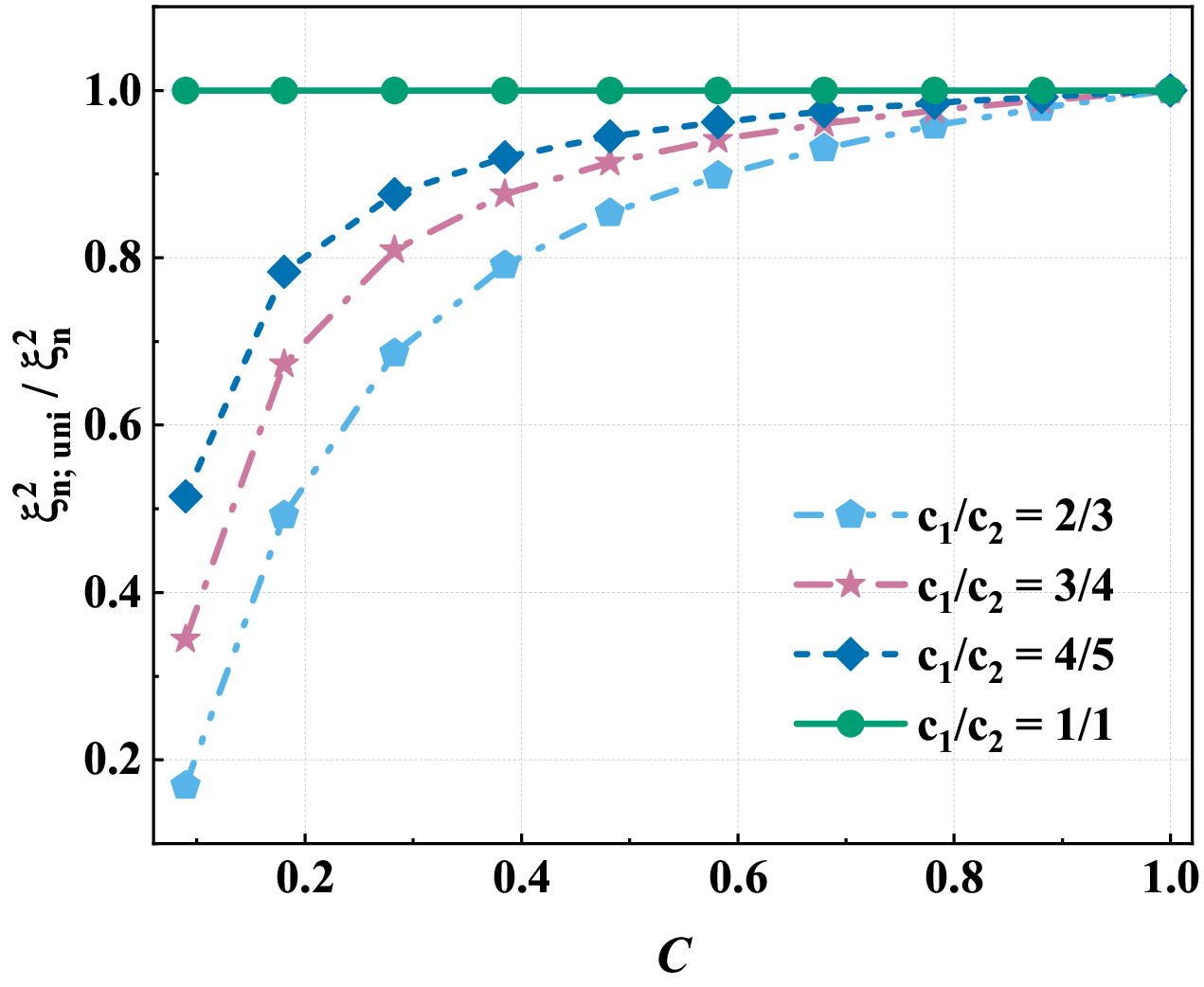}
	\caption{Degradation of the multiparameter spin squeezing under inhomogeneous couplings. For a fixed estimation direction $\mathbf{n}=(1/2,1/2)^\top$, the population fractions are chosen as $\eta_j=|n_j|=1/2$, so that the shot-noise-limited phase sensitivity is minimized. The ratio $\xi_{\mathbf{n};\mathrm{uni}}^2/\xi_\mathbf{n}^2$ is plotted as a function of the curvature $\mathcal{C}$ for a two-ensemble system with representative coupling ratios $c_1/c_2=2/3$, $3/4$, $4/5$, and $1/1$. Here, $\xi_\mathbf{n}^2$ denotes the MSSC for the corresponding coupling configuration, while $\xi_{\mathbf{n};\mathrm{uni}}^2$ is the uniform-coupling value. For all nonuniform cases, the ratio remains below unity, indicating that the cosine-like inhomogeneous coupling increases the MSSC and therefore degrades the attainable quantum gain. The degradation is stronger for larger coupling imbalance and smaller curvature, while all curves approach the uniform-coupling limit as $\mathcal{C} \to 1$.}
	\label{Fig_4}
\end{figure}

To make this degradation quantitative, we further consider a two-ensemble configuration with cosine-like inhomogeneous couplings. For a fixed estimation direction $\mathbf{n}=(1/2,1/2)^\top$, the population fractions are first chosen as $\eta_j=|n_j|=1/2$, which minimize the shot-noise-limited phase sensitivity. Under this choice, the shot-noise contribution is already optimized and fixed, so the effect of coupling inhomogeneity can be isolated directly through the MSSC itself. Figure~\ref{Fig_4} plots the ratio $\xi_{\mathbf{n};\mathrm{uni}}^2/\xi_\mathbf{n}^2$ as a function of the curvature $\mathcal{C}$ for several representative coupling ratios $c_1/c_2$, where $\xi_\mathbf{n}^2$ denotes the MSSC for the corresponding coupling configuration, and $\xi_{\mathbf{n};\mathrm{uni}}^2$ is the uniform-coupling value. For all nonuniform couplings, the ratio remains below unity, implying $\xi_\mathbf{n}^2>\xi_{\mathbf{n};\mathrm{uni}}^2$ and hence a reduced quantum gain. Moreover, the degradation becomes more pronounced with increasing coupling imbalance and decreasing curvature, and all curves approach the uniform-coupling limit as $\mathcal{C}$ converges to unity.

It is also worth emphasizing an alternative comparison strategy. For a given inhomogeneous coupling configuration, one may redistribute the particle numbers among the ensembles so as to optimize the MSSC itself. Because the external drive continuously shifts the steady-state position and hence the curvature of the superradiance potential, this optimized MSSC can in principle be tuned to the same target value for different coupling configurations. In that case, the remaining difference in the total phase sensitivity originates solely from the shot-noise-limited contribution. However, the population fractions required to optimize the MSSC under a given inhomogeneous coupling generally do not satisfy $\eta_j=|n_j|$, and therefore do not minimize the shot-noise-limited phase sensitivity. Consequently, even under this strategy, the uniform-coupling case remains globally optimal, since it is the unique configuration in which the optimization of the MSSC and that of the shot-noise-limited sensitivity can be achieved simultaneously.

Furthermore, from a fundamental physics perspective, the value of this generalized framework extends beyond driven sensing protocols. In a completely uniform open quantum system without coherent driving, permutation symmetry forces the dissipative steady state to be completely factorized (disentangled). However, our inhomogeneous multi-ensemble framework analytically reveals that disparate couplings can dynamically stabilize a non-trivial entangled dark state (a subradiant state) that intrinsically arises from broken permutation symmetry. This ability to rigorously capture both the practical metrological penalties (classical mixedness) and the fundamentally fascinating symmetry-broken entangled states highlights the broad physical validity of our formalism for diverse quantum information tasks.

Before concluding, it is important to briefly discuss the theoretical scope of our model and the practical impact of free-space spontaneous emission. In this work, we have focused on the ideal collective dissipation regime to derive the elegant analytical expressions for the covariance matrix. This approach provides a remarkably clear physical picture of the distributed quantum gain without being obscured by heavy numerical complexities. However, in realistic experimental implementations, single-particle spontaneous emission acts as a local decoherence channel that can compromise the collective interference mechanism. As demonstrated in our previous studies~\cite{Shen2025}, the survival of such collective steady states under local dissipation is rigorously governed by the cooperativity \(\Gamma/\gamma\). As long as the system operates in the high-cooperativity regime, where the collective superradiant coupling rate significantly exceeds the local single-particle dissipation rate, the collective macroscopic dynamics will dominantly overwhelm the local decoherence. Under such conditions, the highly entangled nature of the dark state is sufficiently maintained over an intermediate time scale, ensuring that the distributed metrological sensitivity remains robustly enhanced beyond the SQL.

\section{Conclusions}\label{Sec.4}
In conclusion, we have proposed a driven multi-ensemble superradiant system and derived analytical covariance matrices for its MEDSs, revealing intrinsic inter-ensemble correlations generated by collective dissipation. The resulting covariance matrix is equivalent to the spin-squeezing matrix and exhibits a minimum eigenvalue below unity, unambiguously signaling the presence of multiparameter squeezing. 
We have further introduced the MSSC, which provides a unified variational framework for optimizing quantum-enhanced sensitivity. In particular, for uniform coupling, we have presented a concrete protocol that enables optimal quantum enhancement along arbitrary estimation directions, thereby establishing a direct connection between the geometric structure of the superradiance potential and metrological performance.
Finally, by analyzing finite-size effects and the associated scaling behavior, we have clarified the practical relevance of the proposed scheme beyond the thermodynamic limit. Taken together, our results demonstrate that driven multi-ensemble superradiance constitutes a versatile and intrinsically generated resource for distributed quantum sensing, and offers a promising paradigm for multimode quantum interferometry in driven-dissipative many-body systems.

\section{Acknowledgments}
This work was supported by the National Natural Science Foundation of China (Grant Nos. 12274164, 61875067, and 12375011). We thank Xiuzhu Yang for helpful discussions. 

\section{Data availability}
The data that support the findings of this article is available in Zenodo at \cite{Shen-2025}.

\appendix

\section{Superradiance potential for multi-ensemble superradiance} \label{Appendix A}
Without loss of generality, we choose the torque vector \(\Gamma \boldsymbol{\mathcal{S}}_{\perp}(t) + \mathbf{\Omega}\) along the \(x\)-axis, such that the Bloch vector rotates within the \(y\)-\(z\) plane. Consequently, the time‐dependent angle \(\theta(t)\) obeys the equation
\begin{equation}
	\frac{\upd}{\upd t}\theta(t) =\Omega-\Gamma\langle \hat{\mathcal{S}}^y(t) \rangle,
\end{equation}
where
\begin{equation}
	\langle \hat{\mathcal{S}}^y (t) \rangle=\sum_{j=1}^{M}c_j\langle \hat{S}_j^y (t) \rangle = \sum_{j=1}^{M}c_j\frac{N_j}{2}\sin\left[ c_j\theta(t) \right].
\end{equation}
According to the definition of the superradiance potential~\cite{Sundar2024,Shen2025,Orioli2022}, \(\upd\theta/\upd t = -N\Gamma \upd V(\theta)/\upd\theta\), we obtain its explicit form
\begin{equation}\label{SM_V}
	V(\theta)=-\frac{1}{2}\sum_{j=1}^{M}{\eta_j \cos(c_j \theta)}-\frac{\Omega}{N\Gamma}\theta + V_0,
\end{equation}
with an irrelevant additive constant \(V_0\). The term proportional to \(\theta\) represents the linear potential associated with the driving field, while the remaining cosine terms arise from collective dissipative processes. 

The stationary points of \(V(\theta)\) include local minima, local maxima, and inflection points, defined respectively by \(\upd^2V/\upd\theta^2 >0\), \(<0\), and \(=0\). These correspond to stable, unstable, and neutral equilibria, respectively. In realistic systems, however, quantum fluctuations, even if strongly suppressed, act as perturbations that inevitably destabilize both unstable and neutral equilibria. Hence, only the stable equilibrium (where \(\upd^2V/\upd\theta^2 > 0\)) remains physically meaningful. Based on the explicit form of the superradiance potential given in Eq.~\eqref{SM_V}, the condition for a local minimum is given by
\begin{equation}\label{SM_Omega}
	\Omega(\theta) = \frac{\Gamma}{2}\sum_{j=1}^{M} N_j c_j \sin(c_j \theta),
\end{equation}
which explicitly expresses the balance between coherent driving and collective dissipation.

\section{Dark-state Holstein-Primakoff approximation} \label{Appendix B}
When the system reaches a driven–dissipative equilibrium at the initial time, the first moments of spin operators become effectively time independent. We accordingly rotate each ensemble’s Bloch vector, corresponding to the \(\theta\)-dependent product state, so that it aligns with the ground-state axis of a new coordinate frame. The \(X\) and \(Y\) axes are then redefined within the plane orthogonal to this axis. In this rotated basis, the spin operators can be mapped onto bosonic operators via the Holstein–Primakoff transformation~\cite{Holstein1940,Kurucz2010,Ma2011} as
\begin{equation}
	\left\{
	\begin{aligned}
		&\hat{S}_j^X + \upi \hat{S}_j^Y = \hat{a}_j^\dagger\sqrt{N_j-\hat{a}_j^\dagger \hat{a}_j},\\
		&\hat{S}_j^X - \upi \hat{S}_j^Y = \sqrt{N_j-\hat{a}_j^\dagger \hat{a}_j}\,\hat{a}_j,\\
		&\hat{S}_j^Z = \hat{a}_j^\dagger \hat{a}_j - \frac{N_j}{2},
	\end{aligned}
	\right.
\end{equation}
where \(\hat{a}_j\) is the bosonic annihilation operator of \(j\)th ensemble, satisfying the canonical commutation relation \([\hat{a}_j, \hat{a}_k^\dagger] = \delta_{jk}\). A straightforward calculation confirms that the resulting operators satisfy the standard angular-momentum commutation relations. 

In the rotated coordinate frame, the Bloch vector corresponding to the initial state of each ensemble is regarded as the ground state in the new frame. Throughout the subsequent dynamical evolution, the system evolves in the vicinity of this reference state, in the sense that its first moments remain approximately unchanged. It is therefore justified to assume \(\hat{a}_j^\dagger \hat{a}_j \sim 0\), allowing the Holstein–Primakoff transformation to be linearized as follows
\begin{equation}\label{SM_HP}
	\left\{
	\begin{aligned}
		&\hat{S}_j^X + \upi \hat{S}_j^Y \sim \sqrt{N_j}\hat{a}_j^\dagger,\\
		&\hat{S}_j^X - \upi \hat{S}_j^Y \sim \sqrt{N_j}\hat{a}_j,\\
		&\hat{S}_j^Z \sim - \frac{N_j}{2}.
	\end{aligned}
	\right.
\end{equation}
From the initial state expression in Eq.~\eqref{psi_theta}, we denote the corresponding single-particle state as \(|\phi_{j,0}\rangle=\cos(c_j\theta/2)|g\rangle -\upi\sin(c_j\theta/2)|e\rangle\), with its orthogonal counterpart given by \(\lvert\phi_{j,1}\rangle=-\upi\sin(c_j\theta/2)|g\rangle + \cos(c_j\theta/2)|e\rangle \). These two states thus form the basis of an effective two-level system, satisfying the completeness relation \(|\phi_{j,0}\rangle\langle\phi_{j,0}|+|\phi_{j,1}\rangle\langle\phi_{j,1}|=\mathbbm{1}_2\). In this new basis, referred to as the dark-state basis, we define the collective operators for each ensemble as \(\hat{S}_j^{\alpha\beta}=\sum_{i=1}^{N_j}|\phi_{j,\alpha}\rangle^i\langle \phi_{j,\beta}|\), where \(\alpha, \beta=0,1\). Within this approximation Eq.~\eqref{SM_HP}, the collective operators obey the hierarchy
\begin{equation}\label{SM_S_j^11}
	\hat{S}_j^{11}=\hat{a}_j^\dagger \hat{a}_j \ll \hat{S}_j^{10} \sim \sqrt{N_j}\hat{a}_j^\dagger \ll \hat{S}_j^{00}\sim N_j.
\end{equation}

In the dark-state basis, the lowering operator \(\hat{S}_j^{-}\) of the \(j\)th ensemble can be written as
\begin{align}\label{SM_S_j^-}
	\hat{S}_j^{-}=\sum_{i=1}^{N_j}{\hat{\sigma}_{ge}^{(i)}}&=\sum_{i=1}^{N_j}\sum_{\alpha,\beta=0}^1{\langle \phi_{j,\alpha}^{(i)}\rvert\hat{\sigma}_{ge}^{(i)}\rvert\phi_{j,\beta}^{(i)}\rangle\lvert\phi_{j,\alpha}^{(i)}\rangle\langle \phi_{j,\beta}^{(i)}\rvert} \nonumber 
	\\
	&=\sum_{\alpha,\beta=0}^1 \langle \phi_{j,\alpha}\rvert\hat{\sigma}_{ge}\lvert\phi_{j,\beta}\rangle\sum_{i=1}^{N_j}\lvert\phi_{j,\alpha}^{(i)}\rangle\langle \phi_{j,\beta}^{(i)}\rvert \nonumber
	\\
	&=\sum_{\alpha,\beta=0}^1{(\hat{\sigma}_{ge})_{j,\alpha\beta} \hat{S}_j^{\alpha\beta}},
\end{align}
where \((\hat{\sigma}_{ge})_{j,\alpha\beta}=\langle \phi_{j,\alpha}\rvert\hat{\sigma}_{ge}\lvert\phi_{j,\beta}\rangle\), whose explicit expressions read
\begin{equation}
	\left\{
	\begin{aligned}
		&(\hat{\sigma}_{ge})_{j,00}=-\frac{\upi}{2}\sin\left(c_j\theta\right),\\
		&(\hat{\sigma}_{ge})_{j,01}=\cos^2\left(\frac{c_j\theta}{2}\right),\\
		&(\hat{\sigma}_{ge})_{j,10}=\sin^2\left(\frac{c_j\theta}{2}\right).
	\end{aligned}
	\right.
\end{equation}
Substituting Eq.~\eqref{SM_S_j^11} into Eq.~\eqref{SM_S_j^-} and neglecting the small quantity \(\hat{a}_j^\dagger \hat{a}_j\), we obtain
\begin{align}
	\hat{S}_j^{-}&\sim \sqrt{N_j}\left[\cos^2\left(\frac{c_j}{2}\theta\right)\hat{a}_j+\sin^2\left(\frac{c_j}{2}\theta\right)\hat{a}_j^{\dagger}\right]
	\nonumber
	\\
	&\quad -\upi\frac{N_j}{2}\sin(c_j\theta).
\end{align}
The collective lowering operator \(\hat{D}^{-}\) can thus be expressed as
\begin{align}\label{SM_D-}
	\hat{D}^{-}&= \upi\frac{\Omega}{\Gamma}+\sum_{j=1}^{M}{c_j\hat{S}_j^{-}} \nonumber
	\\
	&\sim \sum_{j=1}^{M}\sqrt{N_j}c_j\left[\cos^2\left(\frac{c_j\theta}{2}\right)\hat{a}_j+\sin^2\left(\frac{c_j\theta}{2}\right)\hat{a}_j^{\dagger}\right] \nonumber
	\\
	&\quad +\frac{\upi}{\Gamma}\left[ \Omega - \frac{\Gamma}{2}\sum_{j=1}^{M}N_j c_j\sin(c_j\theta)\right]\nonumber
	\\
	&=\sum_{j=1}^{M}\sqrt{N_j}c_j\left[\cos^2\left(\frac{c_j\theta}{2}\right)\hat{a}_j+\sin^2\left(\frac{c_j\theta}{2}\right)\hat{a}_j^{\dagger}\right],
\end{align}
where we have employed the drive–dissipation balance, Eq.~\eqref{SM_Omega}, to eliminate the driving term, thereby yielding a multimode Bogoliubov operator. 

\section{Multimode Squeezing and Dark-State Covariance Matrix} \label{Appendix C}
\subsection{Multimode Bogoliubov operators}
Within the multimode Bogoliubov operator in Eq.~\eqref{SM_D-}, all modes are mutually coupled, which complicates the analytical derivation of the dark state’s correlation properties from Eq.~\eqref{drho_dt}. Here, we present a concrete method to reconstruct the multimode Bogoliubov operator in terms of collective mode operators, thereby revealing a structure in which the system couples exclusively to a single collective mode, while the remaining modes decouple and remain frozen in their initial states.

We begin by defining squeezed bosonic annihilation (creation) operators \(\hat{b}_j\) (\(\hat{b}_j^\dagger\)) as linear combinations of the original mode operators \(\hat{a}_j\) and \(\hat{a}_j^\dagger\) appearing in the multimode Bogoliubov operator. For each ensemble \(j\), we rewrite the corresponding Bogoliubov combination as
\begin{align}
	&\cos^2\left(\frac{c_j}{2}\theta\right)\hat{a}_j+\sin^2\left(\frac{c_j}{2}\theta\right)\hat{a}_j^{\dagger} \nonumber
	\\
	&=\sqrt{|\cos(c_j\theta)|}\left[\frac{\cos^2(c_j\theta/2)}{\sqrt{|\cos(c_j\theta)|}}\hat{a}_j+\frac{\sin^2(c_j\theta/2)}{\sqrt{|\cos(c_j\theta)|}}\hat{a}_j^{\dagger}\right] \nonumber
	\\
	&=\left\{  
	\begin{aligned}  
		&\sqrt{\lvert\cos(c_j\theta)\rvert}\left(\hat{a}_j\cosh{r_j}+\hat{a}_j^{\dagger}\sinh{r_j}\right), \\   
		&\sqrt{\lvert\cos(c_j\theta)\rvert}\left(\hat{a}_j\sinh{r_j}+\hat{a}_j^{\dagger}\cosh{r_j}\right), 
	\end{aligned}  
	\right. \nonumber
	\\
	&=\left\{  
	\begin{aligned}  
		&\sqrt{\lvert\cos(c_j\theta)\rvert}\hat{b}_j \quad \mathrm{for} \quad \cos(c_j\theta)>0, \\   
		&\sqrt{\lvert\cos(c_j\theta)\rvert}\hat{b}_j^\dagger \quad \mathrm{for} \quad \cos(c_j\theta)<0,
	\end{aligned}  
	\right. 
\end{align}
where the squeezed bosonic annihilation operators are defined as \(\hat{b}_j = \hat{a}_j\cosh{r_j}+\hat{a}_j^{\dagger}\sinh{r_j}\), with squeezing factors \(r_j\) determined by \(\upe^{-r_j}=\sqrt{\lvert\cos\left(c_j\theta\right)\rvert}\). 
For a given set of angular displacements \(\{c_j\theta\}_M\), we may, without loss of generality, reorder the modes such that the multimode Bogoliubov operator involves \(K\) squeezed annihilation operators and \(M-K\) squeezed creation operators. The collective lowering operator can then be expressed as
\begin{align}\label{SM_D_b}
	\hat{D}^{-} &\sim \sqrt{N}\left(\sum_{j=1}^{K}\sqrt{\eta_j}c_j \upe^{-r_j}\hat{b}_j + \sum_{j=K+1}^{M}\sqrt{\eta_j}c_j \upe^{-r_j}\hat{b}_j^\dagger\right) \nonumber
	\\
	&= \sqrt{N}\left(\sum_{j=1}^{K}\alpha_j \hat{b}_j + \sum_{j=K+1}^{M}\alpha_j \hat{b}_j^\dagger\right),
\end{align}
where \(\alpha_j = \sqrt{\eta_j}c_j \upe^{-r_j}\).

We then construct a collective annihilation operator \(\hat{B}_1\) as a linear combination of the \(K\) squeezed annihilation operators appearing in Eq.~\eqref{SM_D_b}, and similarly define a collective creation operator \(\hat{B}_{K+1}^\dagger\) from the remaining \(M-K\) squeezed creation operators. The procedure proceeds as follows
\begin{align}\label{SM_sum_j}
	&\sum_{j=1}^{K}{\alpha_j \hat{b}_j}+\sum_{j=K+1}^{M}{\alpha_j \hat{b}_j^\dagger} \nonumber
	\\
    &=\left(\sqrt{\sum_{k=1}^{K}{\alpha_k^2}}\right)\sum_{j=1}^{K}{\frac{\alpha_j}{\sqrt{\sum_{k=1}^{K}{\alpha_k^2}}}\hat{b}_j} \nonumber
    \\
    &\quad +\left(\sqrt{\sum_{k=1}^{M-K}{\alpha_{K+k}^2}}\right)\sum_{j=1}^{M-K}{\frac{\alpha_{K+j}}{\sqrt{\sum_{k=1}^{M-K}{\alpha_{K+k}^2}}}\hat{b}_{K+j}^\dagger} \nonumber
	\\
	&=\left(\sqrt{\sum_{k=1}^{K}{\alpha_k^2}}\right)\sum_{j=1}^{K}{u_{j}\hat{b}_j}+\left(\sqrt{\sum_{k=1}^{M-K}\alpha_{K+k}^2}\right)\sum_{j=1}^{M-K}{v_{j}\hat{b}_{K+j}^\dagger} \nonumber
	\\
	&=\left(\sqrt{\sum_{k=1}^{K}{\alpha_k^2}}\right)\hat{B}_1+\left(\sqrt{\sum_{k=1}^{M-K}\alpha_{K+k}^2} \right){\hat{B}_{K+1}^{\dagger}},
\end{align}
where \(\hat{B}_1 = \sum_{j=1}^{K}{u_{j}\hat{b}_j}\) and \(\hat{B}_{K+1}^\dagger = \sum_{j=1}^{M-K}{v_{j}\hat{b}_{K+j}^\dagger}\). The coefficients are given by \(u_{j}=\alpha_j / \sqrt{\sum_{k=1}^{K}{\alpha_k^2}}\) and \(v_{j}=\alpha_{K+j} / \sqrt{\sum_{k=1}^{M-K}{\alpha_{K+k}^2}}\), respectively. To complete the set of collective modes, we introduce two orthogonal matrices
\(\bm{U}\) and \(\bm{V}\), whose first-row elements satisfy \(U_{1j}=u_j\) and \(V_{1j}=v_j\), respectively. These matrices are not uniquely defined; however, this freedom does not affect the physical results. Using \(\bm{U}\) and \(\bm{V}\), the full set of collective \(B\)-mode operators is defined as
\begin{equation}\label{SM_B_j}
	\left\{
	\begin{aligned}
		&\hat{B}_j = \sum_{k=1}^{K}U_{jk}\hat{b}_k \quad \mathrm{for} \quad j=1,2,\ldots,K,\\
		&\hat{B}_{K+l}^\dagger = \sum_{m=1}^{M-K}V_{lm}\hat{b}_{K+m}^\dagger \quad \mathrm{for} \quad l=1,2,\ldots,M-K.
	\end{aligned}
	\right.
\end{equation}
It is straightforward to verify that the collective operators obey the bosonic commutation relations \([\hat{B}_j,\hat{B}_k^\dagger]=\delta_{jk}\).

We now arrive at the final step of diagonalizing the multimode Bogoliubov operator in terms of collective modes. Specifically, we construct a new operator \(\hat{C}_1\) as a linear combination of \(\hat{B}_1\) and \(\hat{B}_{K+1}^\dagger\) via a two-mode squeezing transformation. Crucially, \(\hat{C}_1\) must be an annihilation operator, a requirement enforced by the stability condition \(\mathcal{C}_\theta>0\). The construction proceeds as
\begin{align}
	&\left(\sqrt{\sum_{k=1}^{K}{\alpha_k^2}}\right)\hat{B}_1+\left(\sqrt{\sum_{k=1}^{M-K}\alpha_{K+k}^2} \right){\hat{B}_{K+1}^{\dagger}} \nonumber
	\\
	&=\sqrt{\left(\sum_{k=1}^{K}{\alpha_k^2} - \sum_{k=1}^{M-K}\alpha_{K+k}^2 \right)}\left(\hat{B}_1\cosh{\xi} +\hat{B}_{K+1}^\dagger\sinh{\xi}\right)\nonumber
	\\
	& = \left(\sqrt{\sum_{k=1}^{M}\eta_k c_k^2\cos(c_k\theta)}\right)\hat{C}_1 =\sqrt{\mathcal{C}_{\theta}}\hat{C}_1,
\end{align}
where \(\hat{C}_1=\hat{B}_1\cosh\xi+\hat{B}_{K+1}^\dagger\sinh\xi\), and the two-mode squeezing factor \(\xi\) satisfies
\begin{equation}
	\cosh\xi=\sqrt{\frac{1}{\mathcal{C}_\theta}\sum_{k=1}^{K}\alpha_k^2},
	\qquad
	\sinh\xi=\sqrt{\frac{1}{\mathcal{C}_\theta}\sum_{k=1}^{M-K}\alpha_{K+k}^2}.
\end{equation}
The conjugate mode is given by \(\hat{C}_{K+1}=\hat{B}_1^\dagger\sinh{\xi}+\hat{B}_{K+1}\cosh{\xi}\). while the
remaining \(C\)-mode operators coincide with the previously defined
\(B\)-mode operators. In summary,
\begin{equation}\label{SM_C_j}
	\left\{
	\begin{aligned}
		&\hat{C}_1=\hat{B}_1\cosh\xi+\hat{B}_{K+1}^\dagger\sinh\xi,\\
		&\hat{C}_{K+1}=\hat{B}_1^\dagger\sinh\xi+\hat{B}_{K+1}\cosh\xi,\\
		&\hat{C}_j=\hat{B}_j
		\quad \mathrm{for} \quad j=2,3,\ldots,K,K+2,\ldots,M.
	\end{aligned}
	\right.
\end{equation}
From Eq.~\eqref{SM_D-}, it follows immediately that the collective lowering
operator involves only \(\hat{C}_1\),
\begin{equation}
	\hat{D}^-\sim\sqrt{\mathcal{C}_\theta N}\,\hat{C}_1.
\end{equation}
Substituting this result into the master equation, Eq.~\eqref{drho_dt}, yields
\begin{equation}\label{SM_ddt}
	\frac{\upd\hat{\rho}}{\upd t}
	\sim \mathcal{C}_\theta N\Gamma
	\left(
	\hat{C}_1\hat{\rho}\hat{C}_1^\dagger
	-\frac{1}{2}\hat{\rho}\hat{C}_1^\dagger\hat{C}_1 - \frac{1}{2}\hat{C}_1^\dagger\hat{C}_1\hat{\rho}
	\right).
\end{equation}

\subsection{Collective quadrature operators}
To facilitate the forthcoming analysis of the covariance matrix, we first introduce quadrature operators for each mode as
\begin{equation}
	\left\{
	\begin{aligned}
		&\hat{X}_j^{\beta} = \frac{1}{\sqrt{2}}\left(\hat{\beta}_j + \hat{\beta}_j^\dagger \right),\\
		&\hat{Y}_j^{\beta} = \frac{1}{\sqrt{2}\upi}\left(\hat{\beta}_j - \hat{\beta}_j^\dagger \right),
	\end{aligned}
	\right.
\end{equation}
where \(\beta=a,b,B,C\). These quadrature operators obey the canonical commutation relations \([\hat{X}_j^{\beta},\hat{Y}_k^{\beta}]=\upi\delta_{jk}\). We then collect them into the operator vector
\begin{equation}
	\hat{\mathbf{Q}}^{\beta}=\left(\hat{X}_1^\beta,\ldots,\hat{X}_M^\beta,\hat{Y}_1^\beta,\ldots,\hat{Y}_M^\beta\right)^\top.
\end{equation}

By employing the transformation relations between the squeezed operators \(\hat{b}_j\) and the original operators \(\hat{a}_j\) and \(\hat{a}_j^\dagger\), we obtain
\begin{equation}
	\left\{
	\begin{aligned}
		&\hat{X}_j^{b} = \upe^{r_j}\hat{X}_j^a,\\
		&\hat{Y}_j^{b} = \upe^{-r_j}\hat{Y}_j^a.
	\end{aligned}
	\right.
\end{equation}
It then follows that the quadrature vectors are related by
\begin{equation}
	\hat{\mathbf{Q}}^b=\bm{S}(\mathbf{r})\hat{\mathbf{Q}}^a,
\end{equation}
where the transformation matrix \(\bm{S}(\mathbf{r})\) has the block off-diagonal form
\begin{equation}\label{S_mathbf_r}
	\bm{S}(\mathbf{r})=
	\begin{pmatrix}  
		\begin{matrix}
			\bm{D}            & \bm{0} \\
			\bm{0}       & \bm{D}^{-1}  
		\end{matrix}                            
	\end{pmatrix}_{2M\times 2M},
\end{equation} 
with \(\bm{D}=\mathrm{diag}(e^{r_1},\ldots,e^{r_M})\). The subscript \(2M\times 2M\) indicates the dimension of the matrix.

From Eq.~\eqref{SM_B_j}, the quadrature operators of the \(B\) modes are related to those of the \(b\) modes as
\begin{equation}
	\left\{
	\begin{aligned}
		&\hat{X}_j^{B} = \sum_{k=1}^{K}U_{jk}\hat{X}_k^b,\\
		&\hat{Y}_j^{B} = \sum_{k=1}^{K}U_{jk}\hat{Y}_k^b,
	\end{aligned}
	\right.
\end{equation}
and
\begin{equation}
	\left\{
	\begin{aligned}
		&\hat{X}_{K+j}^{B} = \sum_{k=1}^{M-K}V_{jk}\hat{X}_{K+k}^b,\\
		&\hat{Y}_{K+j}^{B} = \sum_{k=1}^{M-K}V_{jk}\hat{Y}_{K+k}^b.
	\end{aligned}
	\right.
\end{equation}
It then follows that the quadrature vectors are related by
\begin{equation}
	\hat{\mathbf{Q}}^B = \left(\bm{W}\oplus\bm{W}\right)\hat{\mathbf{Q}}^b,
\end{equation}
where \(\bm{W}=\bm{U}\oplus\bm{V}\) is an \(M\times M\) block-diagonal orthogonal matrix.

Finally, by applying Eq.~\eqref{SM_C_j}, we obtain the transformation from the \(B\) modes to the \(C\) modes. In terms of quadrature operators, this transformation reads
\begin{equation}
	\left\{
	\begin{aligned}
		&\hat{X}_{1}^{C} = \hat{X}_1^B\cosh{\xi} + \hat{X}_{K+1}^B\sinh{\xi},\\
		&\hat{Y}_{1}^{C} = \hat{Y}_1^B\cosh{\xi} - \hat{Y}_{K+1}^B\sinh{\xi},
	\end{aligned}
	\right.
\end{equation}
and
\begin{equation}
	\left\{
	\begin{aligned}
		&\hat{X}_{K+1}^{C} = \hat{X}_1^B\sinh{\xi} + \hat{X}_{K+1}^B\cosh{\xi},\\
		&\hat{Y}_{K+1}^{C} = -\hat{Y}_1^B\sinh{\xi} + \hat{Y}_{K+1}^B\cosh{\xi}.
	\end{aligned}
	\right.
\end{equation}
All remaining modes are left unchanged. Collecting these relations, the transformation can be compactly written as
\begin{equation}
	\hat{\mathbf{Q}}^C = \bm{S}(\xi)\hat{\mathbf{Q}}^B,
\end{equation}
where \(\bm{S}(\xi)\) is the symplectic matrix associated with the two-mode squeezing transformation,
\begin{equation}
	\bm{S}(\xi)=
	\begin{pmatrix}  
		\begin{matrix}
			\bm{E}        &\bm{F}  &\bm{0}        &\bm{0}\\
			\bm{F}^\top   &\bm{G}  &\bm{0}        &\bm{0}\\
			\bm{0}        &\bm{0}  &\bm{E}        &-\bm{F}\\
			\bm{0}        &\bm{0}  &-\bm{F}^\top  &\bm{G}
		\end{matrix}                            
	\end{pmatrix}_{2M\times 2M}.
\end{equation}
Here the block matrices are given by
\begin{equation}
	\bm{E}=
	\begin{pmatrix}  
		\begin{matrix}
			\cosh{\xi}        &0       &\cdots        &0\\
			0              &1       &\cdots        &0\\
			\vdots            &\vdots  &\ddots       &\vdots\\
			0              &0       &\cdots        &1
		\end{matrix}                            
	\end{pmatrix}_{K\times K},
\end{equation}
\begin{equation}
	\bm{F}=
	\begin{pmatrix}  
		\begin{matrix}
			\sinh{\xi}        &0       &\cdots        &0\\
			0                 &0       &\cdots        &0\\
			\vdots            &\vdots  &\ddots       &\vdots\\
			0                 &0       &\cdots        &0
		\end{matrix}                            
	\end{pmatrix}_{K\times (M-K)},
\end{equation}
and
\begin{equation}
	\bm{G}=
	\begin{pmatrix}  
		\begin{matrix}
			\cosh{\xi}        &0       &\cdots        &0\\
			0              &1       &\cdots        &0\\
			\vdots            &\vdots  &\ddots       &\vdots\\
			0              &0       &\cdots        &1
		\end{matrix}                            
	\end{pmatrix}_{(M-K) \times (M-K)}.
\end{equation}
Combining this transformation with the orthogonal mode mixing described by \(\bm{W}\oplus\bm{W}\), we define the overall transformation matrix
\begin{equation}
	\bm{R}=\bm{S}(\xi)\left(\bm{W}\oplus\bm{W}\right),
\end{equation}
such that
\begin{equation}
	\hat{\mathbf{Q}}^C = \bm{R}\hat{\mathbf{Q}}^b,
\end{equation}
Explicitly, \(\bm{R}\) takes the block form
\begin{equation}\label{SM_R}
	\bm{R}=
	\begin{pmatrix}  
		\begin{matrix}
			\bm{E}\bm{U}       &\bm{F}\bm{V}  &\bm{0}              &\bm{0}\\
			\bm{F}^\top\bm{U}  &\bm{G}\bm{V}  &\bm{0}              &\bm{0}\\
			\bm{0}             &\bm{0}        &\bm{E}\bm{U}        &-\bm{F}\bm{V}\\
			\bm{0}             &\bm{0}        &-\bm{F}^\top\bm{U}  &\bm{G}\bm{V} 
		\end{matrix}                            
	\end{pmatrix}_{2M \times 2M},
\end{equation}
with inverse
\begin{equation}\label{SM_R^-1}
	\bm{R}^{-1}=
	\begin{pmatrix}  
		\begin{matrix}
			\bm{U}^\top\bm{E}        &-\bm{U}^\top\bm{F}  &\bm{0}                  &\bm{0}\\
			-\bm{V}^\top\bm{F}^\top  &\bm{V}^\top\bm{G}   &\bm{0}                  &\bm{0}\\
			\bm{0}                   &\bm{0}              &\bm{U}^\top\bm{E}       &\bm{U}^\top\bm{F}\\
			\bm{0}                   &\bm{0}              &\bm{V}^\top\bm{F}^\top  &\bm{V}^\top\bm{G} 
		\end{matrix}                            
	\end{pmatrix}_{2M \times 2M}.
\end{equation} 

\subsection{Dark-State Covariance Matrix}
We begin by defining the covariance matrix \(\bm{\Gamma}_Q^\beta\) associated with an arbitrary mode \(\beta\) (\(\beta=a,b,B,C\)), whose elements are
\begin{equation}
	\left(\bm{\Gamma}_Q^\beta\right)_{jk}= \langle \hat{Q}_j^\beta \hat{Q}_k^\beta + \hat{Q}_k^\beta \hat{Q}_j^\beta \rangle.
\end{equation}
Here we have omitted the subtraction of first moments, since all first moments vanish identically for the states considered. Using the linear transformation of quadrature operators, we obtain
\begin{align}
	\left(\bm{\Gamma}_Q^C\right)_{jk} &= \langle \hat{Q}_j^C \hat{Q}_k^C + \hat{Q}_k^C \hat{Q}_j^C \rangle \nonumber
	\\
	 &=\sum_{l,m}R_{jl}R_{km}\langle \hat{Q}_l^b \hat{Q}_m^b + \hat{Q}_m^b \hat{Q}_l^b \rangle \nonumber
	 \\
	 &=\sum_{l,m}R_{jl}R_{km}\left(\bm{\Gamma}_Q^b\right)_{lm} \nonumber
	 \\
	 &=\left(\bm{R}\bm{\Gamma}_Q^b\bm{R}^\top\right)_{jk}.
\end{align}
Since this relation holds for arbitrary indices \(j\) and \(k\), it follows in matrix form that
\begin{equation}
	\bm{\Gamma}_Q^C = \bm{R}\bm{\Gamma}_Q^b\bm{R}^\top.
\end{equation}
Similarly, the covariance matrix of the \(b\) modes is related to that of the original modes by
\begin{equation}\label{SM_Gamma_Q^bb}
	\bm{\Gamma}_Q^b = \bm{S}(\mathbf{r})\bm{\Gamma}_Q^a\bm{S}^\top(\mathbf{r}).
\end{equation}

According to Eq.~\eqref{SM_ddt}, the dissipative dynamics drive the system toward a steady state in which the \(C_1\) mode is cooled to the Gaussian vacuum, while all remaining modes remain frozen in their initial states. Consequently, in the steady-state covariance matrix, all elements associated with the \(C_1\) mode vanish, except for
\(2\langle (\hat{X}_1^C)^2\rangle = 2\langle (\hat{Y}_1^C)^2\rangle = 1\),
whereas all elements unrelated to the \(C_1\) mode retain their initial values. The steady-state covariance matrix in the \(C\)-mode basis therefore takes the form
\begin{equation}
	\bm{\Gamma}_Q^C(\infty) = \bm{\Gamma}_Q^C(0)-\widetilde{\bm{\Gamma}}_Q^C,
\end{equation}
where \(\bm{\Gamma}_Q^C(0)\) denotes the initial covariance matrix of the \(C\) mode, given by
\begin{equation}\label{SM_Gamma^C_jk}
	\Gamma^C_{jk}(0) = \sum_{l,m}R_{jl}R_{km}\Gamma^b_{lm}(0),
\end{equation}
and \(\widetilde{\bm{\Gamma}}_Q^C\) is the correction matrix accounting for the dissipative relaxation of the \(C_1\) mode. Explicitly, this correction matrix has the sparse structure
\begin{widetext}
	\begin{equation}
		\widetilde{\bm{\Gamma}}_Q^C=
		\begin{pmatrix}
			\Gamma_{11}^C(0)-1  &\Gamma_{12}^C(0)  &\cdots &\Gamma_{1,M}^C(0) &\Gamma_{1,M+1}^C(0) &\Gamma_{1,M+2}^C(0) &\cdots & \Gamma_{1,2M}^{C}(0)\\
			\Gamma_{21}^C(0) &0 &\cdots &0 &\Gamma_{2,M+1}^C(0) &0 &\cdots &0  \\
			\vdots & \vdots & \ddots &\vdots &\vdots &\vdots &\ddots &\vdots \\
			\Gamma_{M,1}^C(0) &0 &\cdots &0 &\Gamma_{M,M+1}^C(0) &0 &\cdots &0 \\
			\Gamma_{M+1,1}^C(0) &\Gamma_{M+1,2}^C(0) &\cdots &\Gamma_{M+1,M}^C(0)  & \Gamma_{M+1,M+1}^C(0)-1 &\Gamma_{M+1,M+2}^C(0) &\cdots & \Gamma_{M+1,2M}^C(0)\\
			\Gamma_{M+2,1}^C(0) &0 &\cdots &0 &\Gamma_{M+2,M+1}^C(0) &0 &\cdots &0  \\
			\vdots & \vdots & \ddots &\vdots &\vdots &\vdots &\ddots &\vdots \\
			\Gamma_{2M,1}^C(0) &0 &\cdots &0 &\Gamma_{2M,M+1}^C(0) &0 &\cdots &0 \\
		\end{pmatrix}.
	\end{equation}
\end{widetext}
The transformation from the covariance matrix in the \(C\)-mode basis to that in the \(b\)-mode basis is implemented via the inverse transformation matrix \(\bm{R}^{-1}\), yielding
\begin{equation}
	\bm{\Gamma}_Q^b(\infty) =\bm{\Gamma}_Q^b(0) - \widetilde{\bm{\Gamma}}_Q^b,
\end{equation}
where the correction matrix is given by
\(\widetilde{\bm{\Gamma}}_Q^b = \bm{R}^{-1}\widetilde{\bm{\Gamma}}_Q^C (\bm{R}^{-1})^\top\).
Its matrix elements read
\begin{align}
	\widetilde{\Gamma}_{jk}^b &=\sum_{l,m =1}^{2M}R^{-1}_{jl}R^{-1}_{km}\widetilde{\Gamma}_{lm}^C \nonumber
	\\
	&=\sum_{m=1}^{2M}\left[ R^{-1}_{j1}R^{-1}_{km}\Gamma^C_{1m}(0)+ R^{-1}_{j,M+1}R^{-1}_{km}\Gamma^C_{M+1,m}(0) \right] \nonumber
	\\
	&\quad + \sum_{m=1}^{2M}\left[R^{-1}_{jm}R^{-1}_{k1}\Gamma^C_{m1}(0)+ R^{-1}_{jm}R^{-1}_{k,M+1}\Gamma^C_{m,M+1}(0) \right] \nonumber
	\\
    &\quad - R^{-1}_{j1}R^{-1}_{k1}\Gamma^C_{11}(0) - R^{-1}_{j,M+1}R^{-1}_{k,M+1}\Gamma^C_{M+1,M+1}(0) \nonumber
	\\
	&\quad - R^{-1}_{j,M+1}R^{-1}_{k1}\Gamma^C_{M+1,1}(0) - R^{-1}_{j1}R^{-1}_{k,M+1}\Gamma^C_{1,M+1}(0) \nonumber
	\\
	&\quad - R^{-1}_{j1}R^{-1}_{k1} - R^{-1}_{j,M+1}R^{-1}_{k,M+1}.
\end{align}
Substituting Eq.~\eqref{SM_Gamma^C_jk} into the above expression yields
\begin{align}\label{SM_Gamma_jk^b}
	\widetilde{\Gamma}_{jk}^b &= \sum_{l=1}^{2M}\left[ R^{-1}_{j1}R_{1l}\Gamma^b_{lk}(0)+ R^{-1}_{j,M+1}R_{M+1,l}\Gamma^b_{lk}(0) \right] \nonumber
	\\
	&\quad + \sum_{l=1}^{2M}\left[ R^{-1}_{k1}R_{1l}\Gamma^b_{jl}(0)+ R^{-1}_{k,M+1}R_{M+1,l}\Gamma^b_{jl}(0) \right] \nonumber
	\\
	&\quad -R^{-1}_{j1}R^{-1}_{k1}\sum_{l,m=1}^{2M}R_{1l}R_{1m}\Gamma_{lm}^b(0) \nonumber
	\\
	&\quad -R^{-1}_{j,M+1}R^{-1}_{k,M+1}\sum_{l,m=1}^{2M}R_{M+1,l}R_{M+1,m}\Gamma_{lm}^b(0) \nonumber
	\\
	&\quad -R^{-1}_{j,M+1}R^{-1}_{k1}\sum_{l,m=1}^{2M}R_{M+1,l}R_{1m}\Gamma_{lm}^b(0) \nonumber
	\\
	&\quad -R^{-1}_{j1}R^{-1}_{k,M+1}\sum_{l,m=1}^{2M}R_{1l}R_{M+1,m}\Gamma_{lm}^b(0) \nonumber
	\\
	&\quad - R^{-1}_{j1}R^{-1}_{k1} - R^{-1}_{j,M+1}R^{-1}_{k,M+1}.
\end{align}
For the initial covariance matrix of the \(b\) modes, which is obtained from that of the original \(a\) modes with
\(\bm{\Gamma}_Q^a(0)=\mathbbm{1}_{2M}\), we find the diagonal form
\begin{equation}\label{SM_Gamma_Q^b0}
	\bm{\Gamma}_Q^b(0)=\bm{S}(\mathbf{r})\mathbbm{1}_{2M}\bm{S}^\top(\mathbf{r})=\bm{D}^2\oplus\bm{D}^{-2},
\end{equation}
such that \(\Gamma_{jk}^b(0)=\Gamma_{jj}^b(0)\delta_{jk}\). Exploiting this diagonal structure, Eq.~\eqref{SM_Gamma_jk^b} simplifies to
\begin{align}
	\widetilde{\Gamma}_{jk}^b &= \left( R^{-1}_{j1}R_{1k}+ R^{-1}_{j,M+1}R_{M+1,k}\right)\Gamma^b_{kk}(0) \nonumber
	\\
	&\quad + \left( R^{-1}_{k1}R_{1j}+ R^{-1}_{k,M+1}R_{M+1,j}\right)\Gamma^b_{jj}(0) \nonumber
	\\
	&\quad -R^{-1}_{j1}R^{-1}_{k1}\sum_{m=1}^{2M}R_{1m}^2\Gamma_{mm}^b(0) \nonumber
	\\
	&\quad -R^{-1}_{j,M+1}R^{-1}_{k,M+1}\sum_{m=1}^{2M}R_{M+1,m}^2\Gamma_{mm}^b(0) \nonumber
	\\
	&\quad -R^{-1}_{j,M+1}R^{-1}_{k1}\sum_{m=1}^{2M}R_{M+1,m}R_{1m}\Gamma_{mm}^b(0) \nonumber
	\\
	&\quad -R^{-1}_{j1}R^{-1}_{k,M+1}\sum_{m=1}^{2M}R_{1m}R_{M+1,m}\Gamma_{mm}^b(0) \nonumber
	\\
	&\quad - R^{-1}_{j1}R^{-1}_{k1} - R^{-1}_{j,M+1}R^{-1}_{k,M+1}.
\end{align}
Finally, since both \(\bm{R}\) and \(\bm{R}^{-1}\) are block-diagonal, all mixed \(X\)–\(Y\) elements vanish. The correction matrix of the \(b\) modes therefore assumes the block-diagonal form
\begin{equation}
	\widetilde{\bm{\Gamma}}_Q^b = \widetilde{\bm{\Gamma}}_X^b \oplus \widetilde{\bm{\Gamma}}_Y^b.
\end{equation}

We first consider the case \(j,k \in \{1,2,\ldots,M\}\), which corresponds to the \(X\)-quadrature block of \(\widetilde{\bm{\Gamma}}_Q^b\). The matrix elements read
\begin{align}
	\left(\widetilde{\bm{\Gamma}}_X^b\right)_{jk}&=R^{-1}_{j1}R_{1k}\Gamma_{kk}^b(0) + R^{-1}_{k1}R_{1j}\Gamma_{jj}^b(0) \nonumber
	\\
	&\quad -R^{-1}_{j1}R^{-1}_{k1}\sum_{m=1}^{M}R^2_{1m}\Gamma_{mm}^b(0)-R^{-1}_{j1}R^{-1}_{k1}.
\end{align}
Similarly, for \(j,k \in \{M+1,M+2,\ldots,2M\}\), corresponding to the \(Y\)-quadrature block, we obtain
\begin{align}
	\left(\widetilde{\bm{\Gamma}}_Y^b\right)_{jk} &=R^{-1}_{j,M+1}R_{M+1,k}\Gamma_{kk}^b(0) + R^{-1}_{k,M+1}R_{M+1,j}\Gamma_{jj}^b(0) \nonumber
	\\
	&\quad -R^{-1}_{j,M+1}R_{k,M+1}^{-1}\sum_{m=K+1}^{2M}R^2_{K+1,m}\Gamma_{mm}^b(0) \nonumber
	\\
	&\quad -R^{-1}_{j,M+1}R^{-1}_{k,M+1}.
\end{align}
Inspection of Eqs.~\eqref{SM_R} and \eqref{SM_R^-1} reveals a useful structural relation: the vector formed by the first \(M\) elements of the first row of \(\bm{R}\) coincides with the vector formed by the last \(M\) elements of the \((M+1)\)th column of \(\bm{R}^{-1}\).
We denote this common vector by \(\mathbf{P}\), whose explicit form is
\begin{align}
	\mathbf{P}=&\left(u_1\cosh{\xi}, \ldots, u_K\cosh{\xi}, v_1\sinh{\xi},\ldots,\right. \nonumber
	\\
	&\left. v_{M-K}\sinh{\xi} \right)^\top.
\end{align}
Likewise, the vector composed of the last \(M\) elements of the \((M+1)\)th row of \(\bm{R}\) is identical to the vector formed by the first \(M\) elements of the first column of \(\bm{R}^{-1}\).
We denote this vector by \(\mathbf{T}\), given explicitly by
\begin{align}
	\mathbf{T}=&\left(u_1\cosh{\xi}, \ldots, u_K\cosh{\xi}, -v_1\sinh{\xi}, \ldots,\right.\nonumber
	\\
	&\left. -v_{M-K}\sinh{\xi} \right)^\top.
\end{align}
With these definitions, and using the diagonal form of the initial covariance matrix
\(\bm{\Gamma}_Q^b(0)\) given in Eq.~\eqref{SM_Gamma_Q^b0},
the correction covariance matrices can be expressed in a compact form as
\begin{equation}
	\widetilde{\bm{\Gamma}}_X^b = \mathbf{T}\mathbf{P}^\top \bm{D}^2 + \bm{D}^2\mathbf{P}\mathbf{T}^\top - \mathbf{T}\mathbf{P}^\top\bm{D}^2\mathbf{P}\mathbf{T}^\top -\mathbf{T}\mathbf{T}^\top,
\end{equation}
and 
\begin{equation}
	\widetilde{\bm{\Gamma}}_Y^b = \mathbf{P}\mathbf{T}^\top \bm{D}^{-2} + \bm{D}^{-2}\mathbf{T}\mathbf{P}^\top - \mathbf{P}\mathbf{T}^\top\bm{D}^{-2}\mathbf{T}\mathbf{P}^\top -\mathbf{P}\mathbf{P}^\top.
\end{equation}

Applying the inverse transformation of Eq.~\eqref{SM_Gamma_Q^bb}, we obtain the correction matrix for the original mode \(a\) in the block-diagonal form
\begin{equation}
	\widetilde{\bm{\Gamma}}_Q^a=\widetilde{\bm{\Gamma}}_X^a \oplus 	\widetilde{\bm{\Gamma}}_Y^a.
\end{equation}
The two blocks are related to those of the \(b\) mode via
\begin{align}
	\widetilde{\bm{\Gamma}}_X^a &= \bm{D}^{-1} \widetilde{\bm{\Gamma}}_X^b \bm{D}^{-1} \nonumber
	\\
	&=\bm{D}^{-1}\mathbf{T}\mathbf{P}^\top\bm{D} + \bm{D}\mathbf{P}\mathbf{T}^\top\bm{D}^{-1}\nonumber
	\\ 
	&\quad -\bm{D}^{-1}\mathbf{T}\mathbf{P}^\top\bm{D}^{2}\mathbf{P}\mathbf{T}^\top\bm{D}^{-1} - \bm{D}^{-1}\mathbf{T}\mathbf{T}^\top\bm{D}^{-1},
\end{align} 
and
\begin{align}
	\widetilde{\bm{\Gamma}}_Y^a &= \bm{D} \widetilde{\bm{\Gamma}}_Y^b \bm{D} \nonumber
	\\
	&=\bm{D}\mathbf{P}\mathbf{T}^\top\bm{D}^{-1} + \bm{D}^{-1}\mathbf{T}\mathbf{P}^\top\bm{D}\nonumber
	\\
	& \quad-\bm{D}\mathbf{P}\mathbf{T}^\top\bm{D}^{-2}\mathbf{T}\mathbf{P}^T\bm{D} - \bm{D}\mathbf{P}\mathbf{P}^\top\bm{D}.
\end{align} 
At this stage, it is natural to introduce the vectors \(\mathbf{A}=\bm{D}\mathbf{P}\) and \(\mathbf{B}=\bm{D}^{-1}\mathbf{T}\),
which allow the correction matrices to be written in a compact and symmetric form,
\begin{equation}
	\widetilde{\bm{\Gamma}}_X^a = \mathbf{A}\mathbf{B}^\top + \mathbf{B}\mathbf{A}^\top -\mathbf{B}\mathbf{A}^\top\mathbf{A}\mathbf{B}^T - \mathbf{B}\mathbf{B}^\top,
\end{equation} 
and
\begin{equation}
	\widetilde{\bm{\Gamma}}_Y^a = \mathbf{A}\mathbf{B}^\top + \mathbf{B}\mathbf{A}^\top -\mathbf{A}\mathbf{B}^\top\mathbf{B}\mathbf{A}^T - \mathbf{A}\mathbf{A}^\top.
\end{equation} 
The explicit components of the vectors \(\mathbf{A}\) and \(\mathbf{B}\) read
\begin{align}
	\mathbf{A}=&\left(u_1 \upe^{r_1}\cosh{\xi}, \ldots, u_K \upe^{r_K} \cosh{\xi},\right.  \nonumber
	\\
	&\left. v_1\upe^{r_{K+1}}\sinh{\xi},\ldots,v_{M-K}\upe^{r_M}\sinh{\xi}\right)^\top,
\end{align}
and 
\begin{align}
	\mathbf{B}= &\left(u_1 \upe^{-r_1}\cosh{\xi}, \ldots, u_K \upe^{-r_K} \cosh{\xi},\right. \nonumber
	\\
	&\left. -v_1\upe^{-r_{K+1}}\sinh{\xi},\ldots,-v_{M-K}\upe^{-r_M}\sinh{\xi}\right)^\top.
\end{align}
Expressing all quantities in terms of the physical system parameters \(\{c_j,\eta_j,\theta\}\), we finally obtain
\begin{equation}\label{SM_AB}
	\left\{  
	\begin{aligned}  
		&\mathbf{A} = \sqrt{\frac{1}{\mathcal{C}_{\theta}}}\left(c_1\sqrt{\eta_1}, c_2\sqrt{\eta_2},\ldots, c_M\sqrt{\eta_M}\right)^\top, \\  
		&\mathbf{B} = \sqrt{\frac{1}{\mathcal{C}_{\theta}}}\left(c_1\sqrt{\eta_1}\cos(c_1\theta),\ldots, c_M\sqrt{\eta_M}\cos(c_M\theta)\right)^\top.
	\end{aligned}  
	\right.
\end{equation}
Consequently, the covariance matrix of the original mode \(a\) in the dark state takes the form
\begin{equation}
	\bm{\Gamma}_Q^a(\infty) = \bm{\Gamma}_X^a(\infty)\oplus\bm{\Gamma}_Y^a(\infty),
\end{equation}
where \(\bm{\Gamma}_X^a(\infty) = \mathbbm{1}_M-\widetilde{\bm{\Gamma}}_X^a\) and \(\bm{\Gamma}_Y^a(\infty) = \mathbbm{1}_M-\widetilde{\bm{\Gamma}}_Y^a\). 

\section{Properties and useful formulas for the dark-state covariance matrix} \label{Appendix D}
\subsection{Eigenvalues and eigenvectors}
Before evaluating the eigenvalues of the dark-state covariance matrix \(\bm{\Gamma}_Q^a(\infty)\), we first examine the vectors \(\mathbf{A}\) and \(\mathbf{B}\). Their Euclidean inner product satisfies the identity \(\mathbf{A}^\top\mathbf{B} \equiv 1\), and the norm of \(\mathbf{A}\) is \(\lVert \mathbf{A} \rVert = \mathcal{C}^{-1/2}\), where \(\lVert \mathbf{A} \rVert = \sqrt{\mathbf{A}^\top\mathbf{A}}\). For the correction matrix \(\widetilde{\bm{\Gamma}}_Y^a\), we introduce the shorthand \(\epsilon = 1+\lVert \mathbf{B} \rVert^2\). With this notation, the matrix can be written as
\begin{equation}\label{SM_widetilde_Gamma_X}
	\widetilde{\bm{\Gamma}}_Y^a = \mathbf{A}\mathbf{B}^\top + \mathbf{B}\mathbf{A}^\top -\epsilon\,\mathbf{A}\mathbf{A}^\top.
\end{equation}

Using the Gram--Schmidt orthogonalization procedure, we construct two mutually orthonormal unit vectors \(\mathbf{e}_1\) and \(\mathbf{e}_2\) spanning the subspace generated by \(\mathbf{A}\) and \(\mathbf{B}\). They are defined as
\begin{equation}
	\left\{  
	\begin{aligned}  
		&\mathbf{e}_1 = \frac{\mathbf{A}}{\lVert \mathbf{A} \rVert}, \\  
		&\mathbf{e}_2 = \frac{\mathbf{B}-\frac{\mathbf{A}}{\lVert \mathbf{A} \rVert^2}}{\lVert \mathbf{B}-\frac{\mathbf{A}}{\lVert \mathbf{A} \rVert^2} \rVert}.
	\end{aligned}  
	\right.
\end{equation}
These relations can be inverted to express \(\mathbf{A}\) and \(\mathbf{B}\) in terms of \(\mathbf{e}_1\) and \(\mathbf{e}_2\) as
\begin{equation}\label{SM_A_B}
	\left\{  
	\begin{aligned}  
		&\mathbf{A} = \lVert \mathbf{A} \rVert \mathbf{e}_1, \\  
		&\mathbf{B} = \frac{\mathbf{e}_1}{\lVert \mathbf{A} \rVert}+\frac{\sqrt{\lVert\mathbf{A}\rVert^2\lVert\mathbf{B}\rVert^2-1}}{\lVert \mathbf{A} \rVert}\mathbf{e}_2.
	\end{aligned}  
	\right.
\end{equation}
We further define the angle \(\phi\) between \(\mathbf{A}\) and \(\mathbf{B}\) via \(\cos\phi = \mathbf{A}^\top\mathbf{B}/(\lVert \mathbf{A} \rVert \lVert \mathbf{B} \rVert)\). Since \(\mathbf{A}^\top\mathbf{B}=1\) and \(\lVert \mathbf{A} \rVert \ge \lVert \mathbf{B} \rVert\), it follows that
\begin{equation}
	\lVert\mathbf{A} \rVert^2 \ge \sec\phi \ge \lVert\mathbf{B} \rVert^2 .
\end{equation}

By substituting Eqs.~\eqref{SM_A_B} into Eq.~\eqref{SM_widetilde_Gamma_X}, we obtain
\begin{align}
	\widetilde{\bm{\Gamma}}_Y^a &= \left( 2-\epsilon\lVert \mathbf{A} \rVert^2 \right)\mathbf{e}_1\mathbf{e}_1^\top \nonumber
	\\
	&\quad + \sqrt{\lVert \mathbf{A} \rVert^2\lVert \mathbf{B} \rVert^2 -1}\left(\mathbf{e}_1\mathbf{e}_2^\top + \mathbf{e}_2\mathbf{e}_1^\top \right).
\end{align}
This matrix has support only in the two-dimensional subspace spanned by \(\{\mathbf{e}_1,\mathbf{e}_2\}\), and can therefore be diagonalized analytically. Its two nonzero eigenvalues are
\begin{equation}\label{SM_lambda_X^pm}
	\lambda_Y^{\pm} = \frac{1}{2}\left( b_Y \pm \sqrt{b_Y^2 + 4 c^2} \right),
\end{equation}
where \(b_Y = 2-\epsilon \lVert\mathbf{A}\rVert^2\) and \(c^2 = \lVert \mathbf{A} \rVert^2\lVert \mathbf{B} \rVert^2 - 1\). The corresponding normalized eigenvectors are
\begin{equation}
	\left\{  
	\begin{aligned}  
		&\mathbf{v}_Y^+ = \cos\!\left(\frac{\alpha_y}{2}\right)\mathbf{e}_1 + \sin\!\left(\frac{\alpha_y}{2}\right)\mathbf{e}_2, \\  
		&\mathbf{v}_Y^- = -\sin\!\left(\frac{\alpha_y}{2}\right)\mathbf{e}_1 + \cos\!\left(\frac{\alpha_y}{2}\right)\mathbf{e}_2,
	\end{aligned}  
	\right.
\end{equation} 
where \(\alpha_y \in [\pi/2, \pi]\) is defined by \(\tan \alpha_y=\sin(2\phi)/\left(\cos(2\phi)-\lVert \mathbf{B} \rVert^{-2}\right)\).

Proceeding analogously, we obtain the eigenvalues of \(\widetilde{\bm{\Gamma}}_X^a\) as
\begin{equation}\label{SM_lambda_Y^pm}
	\lambda_X^{\pm} = \frac{1}{2}\left( b_X \pm \sqrt{b_X^2 + 4 c^2} \right),
\end{equation}
where \(b_X=2-\varepsilon\lVert \mathbf{B} \rVert^2\), with \(\varepsilon=1+\lVert \mathbf{A} \rVert^2\). As in the \(Y\)-quadrature case, the matrix \(\widetilde{\bm{\Gamma}}_X^a\) has support only in the two-dimensional subspace spanned by \(\{\mathbf{e}_1,\mathbf{e}_2\}\). The corresponding normalized eigenvectors are
\begin{equation}
	\left\{  
	\begin{aligned}  
		&\mathbf{v}_X^+ = \cos\!\left(\frac{\alpha_x}{2}\right)\mathbf{e}_1 - \sin\!\left(\frac{\alpha_x}{2}\right)\mathbf{e}_2, \\  
		&\mathbf{v}_X^- = \sin\!\left(\frac{\alpha_x}{2}\right)\mathbf{e}_1 + \cos\!\left(\frac{\alpha_x}{2}\right)\mathbf{e}_2,
	\end{aligned}  
	\right.
\end{equation} 
where \(\alpha_x \in [0, \pi/2]\) is defined by \(\tan \alpha_x=\sin(2\phi)/\left(\lVert \mathbf{A} \rVert^{2}-\cos(2\phi)\right)\).

In summary, the asymptotic covariance matrix \(\bm{\Gamma}_X^a(\infty)\) exhibits two nontrivial eigenvalues, given by \(1-\lambda_X^{\pm}\), with the corresponding eigenvectors \(\mathbf{v}_X^{\pm}\). Likewise, \(\bm{\Gamma}_Y^a(\infty)\) features two eigenvalues that deviate from unity, \(1-\lambda_Y^{\pm}\), associated with the eigenvectors \(\mathbf{v}_Y^{\pm}\).

\subsection{Spectrum visualization and asymptotic scaling of the lowest eigenpair of \(\mathbf{\Gamma}_Q^a(\infty)\)}
From the explicit expressions of \(\lambda_X^{\pm}\) and \(\lambda_Y^{\pm}\) [Eqs.~\eqref{SM_lambda_X^pm} and \eqref{SM_lambda_Y^pm}], one finds that they share the same functional form, differing only in the coefficients \(b_X\) and \(b_Y\). Since \(\lVert \mathbf{A} \rVert \geq \lVert \mathbf{B} \rVert\), it follows immediately that \(b_X \geq b_Y\). To identify the minimum eigenvalue of \(\bm{\Gamma}_Q^a(\infty)\), it is therefore sufficient to compare \(1-\lambda_X^+\) and \(1-\lambda_Y^+\). Noting that the function \(f(x)=x+\sqrt{x^2+4c^2}\) is nonnegative and monotonically increasing in \(x\), we conclude that \(\lambda_X^+ \geq \lambda_Y^+\). Hence, the minimum eigenvalue of \(\bm{\Gamma}_Q^a(\infty)\) is given by \(\lambda_Q^{\mathrm{min}} = 1 - \lambda_X^+\).

\begin{figure}[t]
	\centering
	\includegraphics[scale=0.48]{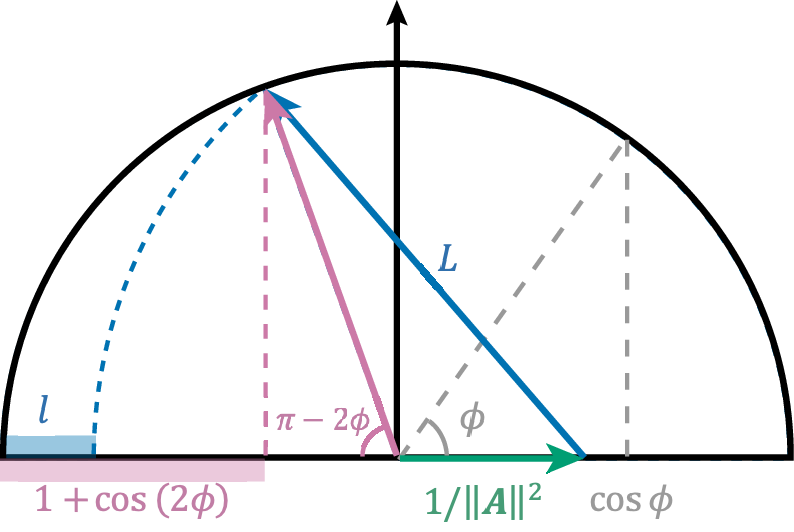}
	\caption{Graphical representation of the minimum eigenvalue of \(\bm{\Gamma}_Q^a(\infty)\).}
	\label{Fig_5}
\end{figure}

Since \(\mathbf{A}^\top\mathbf{B}\equiv 1\), the quantities \(\lVert\mathbf{A}\rVert\), \(\lVert\mathbf{B}\rVert\), and \(\cos\phi\) are not independent. It is therefore convenient to eliminate \(\lVert\mathbf{B}\rVert\) in favor of \(\lVert\mathbf{A}\rVert\) and \(\cos\phi\), which yields \(\lVert \mathbf{B} \rVert = 1 / \left( \lVert \mathbf{A} \rVert \cos\phi \right)\). In this parametrization, \(b_X\) and \(c\) become
\begin{equation}
	b_X = \frac{\cos(2\phi)-1/\lVert \mathbf{A} \rVert^2}{\cos^2\phi},
	\qquad
	c=\tan\phi.
\end{equation}
Substituting these expressions into \(\lambda_Q^{\mathrm{min}}\) leads to
\begin{widetext}
	\begin{equation}\label{SM_lambda_Q^min}
		\lambda_Q^{\mathrm{min}}= \frac{1}{1+\cos(2\phi)}\left[1+\lVert \mathbf{A} \rVert^{-2} - \sqrt{1+\lVert \mathbf{A} \rVert^{-4}-2\lVert \mathbf{A} \rVert^{-2}\cos(2\phi)}\right].
	\end{equation}
\end{widetext}
As illustrated in Fig.~\ref{Fig_5}, we consider a semicircle of unit radius, along which the minimum eigenvalue of \(\bm{\Gamma}_Q^a(\infty)\) is visualized. The constraint \(\mathbf{A}^\top\mathbf{B}=1\) implies \(\phi<\pi/2\). By the law of cosines, one finds that the square-root term in Eq.~\eqref{SM_lambda_Q^min} corresponds exactly to the length \(L\) indicated in Fig.~\ref{Fig_5}.
Taking \(L\) as the radius and its initial point as the center, a circle is drawn that intersects the horizontal axis. The distance between the left endpoint of the semicircle and the intersection point is denoted by \(l\), corresponding to the bracketed term in Eq.~\eqref{SM_lambda_Q^min} (shown as the blue segment). The denominator \(1+\cos(2\phi)\) is represented by the red segment. Hence, \(\lambda_Q^{\mathrm{min}}\) is given by the ratio of the blue and red segments, which is manifestly smaller than unity.

As \(\lVert \mathbf{A} \rVert\) increases, the initial point of the segment \(L\) approaches the center of the semicircle, and the corresponding length \(l\) decreases accordingly. Consequently, a sufficiently large norm of \(\mathbf{A}\) leads to a sufficiently small value of \(\lambda_Q^{\mathrm{min}}\). This geometric relation establishes a direct correspondence between the curvature of the superradiance potential and the spectrum of the dark-state covariance matrix: dark states associated with smaller potential curvature
exhibit smaller minimum eigenvalues \(\lambda_Q^{\mathrm{min}}\).

In the regime \(\lVert\mathbf{A}\rVert \gg 1\), we obtain the asymptotic expansion
\begin{align}
	\lambda_Q^{\mathrm{min}}&\sim\frac{1}{1+\cos(2\phi)}\left[ \frac{1+\cos(2\phi)}{\lVert \mathbf{A} \rVert^2}-\frac{1}{2\lVert \mathbf{A} \rVert^4} \right] \nonumber
	\\
	&= \lVert \mathbf{A} \rVert^{-2} + \mathcal{O}\left( \lVert \mathbf{A} \rVert^{-4} \right).
\end{align}
Meanwhile,
\begin{align}
	\tan{\alpha_x}=\frac{\sin(2\phi)}{\lVert \mathbf{A} \rVert^2-\cos(2\phi)}\sim \frac{\sin(2\phi)}{\lVert \mathbf{A} \rVert^2}\sim \alpha_x,
\end{align} 
so that
\begin{equation}\label{SM_cos_sin}
	\left\{  
	\begin{aligned}  
		&\cos\!\left(\frac{\alpha_x}{2} \right)\sim 1-\frac{1}{2}\left(\frac{\alpha_x}{2}\right)^2\sim 1 - \frac{\sin^2(2\phi)}{8\lVert \mathbf{A} \rVert^4}, \\  
		&\sin\!\left(\frac{\alpha_x}{2} \right)\sim \frac{\alpha_x}{2}\sim \frac{\sin(2\phi)}{2\lVert \mathbf{A} \rVert^2}.
	\end{aligned}  
	\right.
\end{equation} 
Therefore, in the large-\(\lVert\mathbf{A}\rVert\) limit, the eigenvector \(\mathbf{v}_X^+\) associated with the minimum eigenvalue
becomes asymptotically aligned with \(\mathbf{A}\).

\subsection{The Matrix Difference \(\mathbf{\Gamma}_X^a(\infty)-[\mathbf{\Gamma}_Y^{a}(\infty)]^{-1}\)}
Using the parametrization in Eq.~\eqref{SM_A_B}, the covariance matrices \(\bm{\Gamma}_X^a(\infty)\) and \(\bm{\Gamma}_Y^a(\infty)\) can be written as
\begin{align}
	\bm{\Gamma}_X^a(\infty) &= \frac{1}{\lVert \mathbf{A} \rVert^2}\left[ \mathbf{e}_1\mathbf{e}_1^\top + \left( \varepsilon c^2 + \lVert \mathbf{A} \rVert^2 \right)\mathbf{e}_2\mathbf{e}_2^\top \right.   \nonumber
	\\
	&\quad + \left. c\left(\mathbf{e}_1\mathbf{e}_2^\top + \mathbf{e}_2\mathbf{e}_1^\top\right) \right] + \sum_{m=3}^{M}\mathbf{e}_m\mathbf{e}_m^\top,
\end{align} 
and
\begin{align}
	\bm{\Gamma}_Y^a(\infty) &= \left(1-b_Y\right)\mathbf{e}_1\mathbf{e}_1^\top + \mathbf{e}_2\mathbf{e}_2^\top - c\left(\mathbf{e}_1\mathbf{e}_2^\top + \mathbf{e}_2\mathbf{e}_1^\top\right) \nonumber
	\\
	&\quad + \sum_{m=3}^{M}\mathbf{e}_m\mathbf{e}_m^\top,
\end{align} 
where \(\{\mathbf{e}_3,\ldots,\mathbf{e}_M\}\) span the subspace orthogonal to that generated by
\(\mathbf{A}\) and \(\mathbf{B}\). From \(\bm{\Gamma}_Y^a(\infty)\), its inverse is readily obtained as
\begin{align}
	[\bm{\Gamma}_Y^{a}(\infty)]^{-1} &= \frac{1}{\lVert \mathbf{A} \rVert^2}\left[ \mathbf{e}_1\mathbf{e}_1^\top + \left( 1-b_Y \right)\mathbf{e}_2\mathbf{e}_2^\top \right. \nonumber
	\\
	&\quad \left. + c\left(\mathbf{e}_1\mathbf{e}_2^\top + \mathbf{e}_2\mathbf{e}_1^\top\right) \right] + \sum_{m=3}^{M}\mathbf{e}_m\mathbf{e}_m^\top.
\end{align} 
Therefore, the difference between the two matrices is given by
\begin{equation}
	\bm{\Gamma}_X^a(\infty)-[\bm{\Gamma}_Y^{a}(\infty)]^{-1}= c^2 \mathbf{e}_2\mathbf{e}_2^\top.
\end{equation} 
Expressed in terms of the angle \(\phi\) between \(\mathbf{A}\) and \(\mathbf{B}\), this relation becomes
\begin{equation}\label{SM_Ya_Xa}
	\bm{\Gamma}_X^a(\infty)-[\bm{\Gamma}_Y^{a}(\infty)]^{-1}= \left( \tan^2\phi \right) \mathbf{e}_2\mathbf{e}_2^\top.
\end{equation} 
This result demonstrates that the closer \(\mathbf{A}\) and \(\mathbf{B}\) are to being parallel, the smaller the deviation between \(\bm{\Gamma}_X^a(\infty)\) and \([\bm{\Gamma}_Y^a(\infty)]^{-1}\). In particular, when \(\mathbf{A}\parallel\mathbf{B}\),
\begin{equation}
	\bm{\Gamma}_X^a(\infty) = [\bm{\Gamma}_Y^{a}(\infty)]^{-1}.
\end{equation} 

\section{The relation between the multiparameter squeezing coefficient and the squeezing matrix}\label{Appendix E}
In multiparameter quantum metrology, one often aims to estimate a linear combination of unknown parameters,
\begin{equation}
	\varphi_{\mathbf n}:=\mathbf n^\top \boldsymbol{\varphi} = \sum_{j=1}^{M} n_j \varphi_j ,
\end{equation}
where \(\boldsymbol{\varphi} = (\varphi_1,\varphi_2,\ldots,\varphi_M)^\top\) denotes the vector of unknown parameters and
\(\mathbf n = (n_1,n_2,\ldots,n_M)^\top \in \mathbb{R}^M\) specifies the estimation direction, subject to the normalization
\(\sum_{j=1}^{M} |n_j| = 1\). Typical examples include the estimation of a field gradient, \(\mathbf n=(1/2,-1/2)^\top\), and of the average field, \(\mathbf n=(1/2,1/2)^\top\). The parameters are imprinted onto the probe state \(\hat{\rho}\) via the unitary evolution \(\hat{U}(\boldsymbol{\varphi}) = \exp(-\upi\boldsymbol{\varphi}^\top \hat{\mathbf{G}}) = \exp(-\upi\sum_{j=1}^{M} \varphi_j\hat{G}_j)\), where \(\hat{\mathbf{G}} = (\hat{G}_1, \hat{G}_2,\ldots,\hat{G}_M)^\top\) is a vector of generators, which are not required to commute. After the parameter encoding, observables \(\hat{\mathbf{X}} = (\hat{X}_1,\hat{X}_2,\ldots,\hat{X}_M)^\top\) are measured, and the experiment is repeated \(\mu\) times using the same output state \(\hat{\rho}(\boldsymbol{\varphi}) = \hat{U}(\boldsymbol{\varphi})\hat{\rho}\hat{U}^\dagger(\boldsymbol{\varphi})\). The individual parameters \(\varphi_j\) are inferred from estimators \(\varphi_{\mathrm{est},j}\), which are functions of the measurement outcomes. Accordingly, the estimator for \(\varphi_{\mathbf n}\) is given by \(\mathbf n^\top \boldsymbol{\varphi}_{\mathrm{est}}\), with variance
\begin{equation}
	(\Delta \mathbf n^\top \boldsymbol{\varphi}_{\mathrm{est}})^2
	= \mathbf{n}^\top \bm{\Sigma}\, \mathbf{n},
\end{equation}
where \(\bm{\Sigma}\) is the covariance matrix of the estimators, with elements \(\Sigma_{jk}=\mathrm{Cov}(\varphi_{\mathrm{est},j},\varphi_{\mathrm{est},k})\).

In the central limit, the covariance matrix \(\bm{\Sigma}\) of the estimators is given by~\cite{Gessner2020}
\begin{equation}\label{SM_Sigma}
	\bm{\Sigma} = \left( \mu \mathcal{M}_{\boldsymbol{\varphi}}[\hat{\mathbf{G}},\hat{\mathbf{X}}] \right)^{-1}.
\end{equation}
Here, the moment matrix reads
\begin{equation}
	\mathcal{M}_{\boldsymbol{\varphi}}[\hat{\mathbf{G}},\hat{\mathbf{X}}] = \bm{C}^\top[\hat{\mathbf{G}},\hat{\mathbf{X}}] \bm{\Gamma}^{-1}[\hat{\mathbf{X}}] \bm{C}[\hat{\mathbf{G}},\hat{\mathbf{X}}],
\end{equation}
where the commutator matrix has elements \(C_{jk}[\hat{\mathbf{G}},\hat{\mathbf{X}}] = -i \langle [\hat{X}_j,\hat{G}_k] \rangle\), and the covariance matrix of the measurement observables is defined as \(\Gamma_{jk}[\hat{\mathbf{X}}] = \frac{1}{2}\langle \hat{X}_j\hat{X}_k + \hat{X}_k\hat{X}_j \rangle - \langle \hat{X}_j \rangle \langle \hat{X}_k \rangle\). Equation~\eqref{SM_Sigma} constitutes a multiparameter generalization of the single-parameter error–propagation formula. The moment matrix satisfies the hierarchy~\cite{Gessner2018,Liu2020}
\begin{equation}\label{SM_MFG}
	\mathcal{M}_{\boldsymbol{\varphi}}[\hat{\mathbf{G}},\hat{\mathbf{X}}] \leq \bm{F}_Q[\hat{\mathbf{G}}] \leq 4 \bm{\Gamma}[\hat{\mathbf{G}}],
\end{equation}
where \(\bm{F}_Q[\hat{\mathbf{G}}]\) denotes the QFIM, which is symmetric and positive semidefinite, and is determined by both the generators \(\hat{\mathbf{G}}\) and the probe state. Here, \(\bm{\Gamma}[\hat{\mathbf{G}}]\) is the covariance matrix associated with the generators. The upper bound is saturated for pure probe states.

In the multiparameter scenario, the squeezing matrix, introdued by Gessner \emph{et al.}~\cite{Gessner2020}, is defined through a comparison between the moment-based sensitivity matrix \(\bm{\Sigma}\) in Eq.~\eqref{SM_Sigma} and the multiparameter shot-noise limit \(\bm{\Sigma}_{\mathrm{SN}} = \left( \mu\, \bm{F}_{\mathrm{SN}}[\hat{\mathbf{G}}] \right)^{-1}\), which represents the ultimate sensitivity achievable with separable probe states. Here, \(\bm{F}_{\mathrm{SN}}[\hat{\mathbf{G}}]:=\underset{\hat{\rho}_{\mathrm{sep}}}{\max}\,\bm{F}_Q[\hat{\rho}_{\mathrm{sep}},\hat{\mathbf{G}}]\) denotes the maximal QFIM attainable using an optimally chosen separable probe state. For parameter imprinting generated by \(\hat{\mathbf{G}}\) and measurements of observables \(\hat{\mathbf{X}}\), the multiparameter squeezing matrix is defined as
\begin{equation}\label{SM_Xi2_GX}
	\bm{\Xi}^2[\hat{\mathbf{G}},\hat{\mathbf{X}}] := \bm{F}^{1/2}_{\mathrm{SN}}[\hat{\mathbf{G}}] \mathcal{M}^{-1}_{\boldsymbol{\varphi}}[\hat{\mathbf{G}},\hat{\mathbf{X}}]\bm{F}^{1/2}_{\mathrm{SN}}[\hat{\mathbf{G}}].
\end{equation}
Any quantum state satisfying \(\bm{\Xi}^2[\hat{\mathbf{G}},\hat{\mathbf{X}}] \geq \mathbbm{1}_M\) is limited to shot-noise–level sensitivity in multiparameter estimation. A violation of this matrix inequality therefore certifies the presence of multiparameter squeezing with respect to the generators \(\hat{\mathbf{G}}\) and the measurement observables \(\hat{\mathbf{X}}\).

For a given estimation direction \(\mathbf{n}\), and a fixed configuration of generators and measurement observables \(\{\hat{\mathbf{G}},\hat{\mathbf{X}}\}\), we define the \emph{multiparameter squeezing coefficient}, the natural analog of the squeezing coefficient in single-parameter estimation, as
\begin{equation}\label{SM_xi_n^2}
	\xi_{\mathbf{n}}^{2}[\hat{\mathbf{G}},\hat{\mathbf{X}}]:=\frac{\left(\Delta \varphi_\mathbf{n}\right)^2}{\left(\Delta \varphi_\mathbf{n}\right)^2_\mathrm{SN}}=\frac{\mathbf{n}^\top \mathcal{M}^{-1}_{\boldsymbol{\varphi}}[\hat{\mathbf{G}},\hat{\mathbf{X}}]\mathbf{n}}{\mathbf{n}^\top \bm{F}^{-1}_\mathrm{SN}[\hat{\mathbf{G}}]\,\mathbf{n}}.
\end{equation}
A value \(\xi_{\mathbf{n}}^{2}<1\) indicates that the corresponding sensitivity cannot be achieved by any separable probe state within the given configuration, but instead necessarily originates from quantum entanglement, enabling sensitivity beyond the SQL. In Eq.~\eqref{SM_xi_n^2}, the shot-noise–limited sensitivity \(\left( \Delta \varphi_{\mathbf{n}} \right)^2_\mathrm{SN}\) represents the optimal classical sensitivity attainable for the specified generators \(\hat{\mathbf{G}}\). 

We now elucidate the relation between the multiparameter squeezing coefficient and the squeezing matrix. Starting from Eq.~\eqref{SM_xi_n^2}, and introducing the transformed vector \(\mathbf{m} = \bm{F}^{-1/2}_\mathrm{SN}[\hat{\mathbf{G}}] \mathbf{n}\),
the multiparameter squeezing coefficient can be rewritten as
\begin{equation}\label{SM_RQ}
	\xi_{\mathbf{n}}^{2}[\hat{\mathbf{G}},\hat{\mathbf{X}}] = \frac{\mathbf{m}^\top \bm{\Xi}^2[\hat{\mathbf{G}},\hat{\mathbf{X}}]\mathbf{m}}{\mathbf{m}^\top\mathbf{m}}\equiv\mathcal{R}[\bm{\Xi}^2,\mathbf{m}],
\end{equation}
which is the \emph{Rayleigh quotient} of the squeezing matrix \(\bm{\Xi}^2[\hat{\mathbf{G}},\hat{\mathbf{X}}]\) with respect to \(\mathbf{m}\). It then follows directly from the properties of the Rayleigh quotient that
\begin{equation}
	\min_{\mathbf{n \in \mathbb{R}^M}}\xi^2_\mathbf{n}[\hat{\mathbf{G}}, \hat{\mathbf{X}}] = \lambda_{\mathrm{min}}[\bm{\Xi}^2],
\end{equation}
where \(\lambda_{\mathrm{min}}[\bm{\Xi}^2]\) denotes the minimum eigenvalue of the squeezing matrix. This result shows that, for a fixed choice of generators \(\hat{\mathbf{G}}\), measurement observables \(\hat{\mathbf{X}}\), and probe state \(\hat{\rho}\), a multimode quantum interferometer possesses a characteristic estimation direction along which the available quantum resources are optimally exploited. Equivalently, the minimization of \(\xi_{\mathbf{n}}^{2}\) can be viewed as a variational principle in which the system parameters are tuned such that the eigenvector associated with the minimum eigenvalue of the squeezing matrix aligns with the transformed direction \(\mathbf{m}\), thereby yielding optimal quantum gain.

In this work, the squeezing matrix takes the explicit form \(\mathbbm{1}_M - \widetilde{\bm{\Gamma}}_X^a\) . In the limit where its minimum eigenvalue approaches zero, the corresponding eigenvector converges to \(\mathbf{e}_1\), as shown in Eq.~\eqref{SM_cos_sin}. Consequently, optimal quantum gain is achieved when the transformed vector \(\mathbf{m}\) is aligned with \(\mathbf{A}\), namely,
\begin{equation}\label{SM_n_c}
	\frac{\lvert n_1 \rvert}{\lvert c_1 \rvert \eta_1} = \frac{\lvert n_2 \rvert}{\lvert c_2 \rvert \eta_2} = \cdots \frac{\lvert n_M \rvert}{\lvert c_M \rvert \eta_M}.
\end{equation}
The global parameter sensitivity can be expressed in terms of the multiparameter squeezing coefficient as
\begin{equation}
	\left(\Delta \varphi_\mathbf{n}\right)^2 = \xi^2_{\mathbf{n}} \left(\Delta \varphi_\mathbf{n}\right)^2_{\mathrm{SN}}.
\end{equation}
The atom numbers are first distributed among the individual ensembles so as to optimize the shot-noise–limited sensitivity,
\begin{equation}
	\left(\Delta\varphi_\mathbf{n}\right)^2_{\mathrm{SN}} = \left(\Delta\varphi_\mathbf{n}\right)^2_{\mathrm{SN},\mathrm{opt}} = \frac{1}{N},
\end{equation}
which is achieved for population fractions \(\eta_j = \lvert n_j \rvert\). It then follows directly from Eq.~\eqref{SM_n_c} that
\(\lvert c_1 \rvert = \lvert c_2 \rvert = \cdots = \lvert c_M \rvert\), demonstrating that this strategy yields the optimal global parameter sensitivity.

It is worth emphasizing that, although this strategy is derived in the limit where the multiparameter squeezing coefficient approaches zero, it remains valid for arbitrary squeezing strengths. Under this strategy, the eigenvector associated with the minimum eigenvalue of the squeezing matrix is strictly given by \(\mathbf{e}_1\), and the correction matrix \(\widetilde{\bm{\Gamma}}_X^a\) reduces from the two-dimensional subspace spanned by \(\mathbf{A}\) and \(\mathbf{B}\) to the one-dimensional subspace defined solely by \(\mathbf{A}\).

In this uniform coupling case (\(\lvert c_1 \rvert = \lvert c_2 \rvert = \cdots = \lvert c_M \rvert\)), the squeezing matrix simplifies significantly and assumes a highly symmetric explicit form,
\begin{widetext}
	\begin{equation}
		\mathbbm{1}_M - \widetilde{\bm{\Gamma}}_X^a =
		\begin{pmatrix}
			1-(1-\mathcal{C})\eta_1 & \mathrm{sgn}(c_1c_2)(\mathcal{C}-1)\sqrt{\eta_1\eta_2} & \cdots &  \mathrm{sgn}(c_1c_M)(\mathcal{C}-1)\sqrt{\eta_1\eta_M}\\
			\mathrm{sgn}(c_1c_2)(\mathcal{C}-1)\sqrt{\eta_1\eta_2} & 1-(1-\mathcal{C})\eta_2 & \cdots & \mathrm{sgn}(c_2c_M)(\mathcal{C}-1)\sqrt{\eta_2\eta_M} \\
			\vdots & \vdots & \ddots & \vdots \\
			\mathrm{sgn}(c_1c_M)(\mathcal{C}-1)\sqrt{\eta_1\eta_M} & \mathrm{sgn}(c_2c_M)(\mathcal{C}-1)\sqrt{\eta_2\eta_M} & \cdots & 1-(1-\mathcal{C})\eta_M
		\end{pmatrix},
	\end{equation}
\end{widetext}
where the spin-squeezing coefficient of the \(j\)th ensemble, namely the \(j\)th diagonal element of the squeezing matrix, is \(1-(1-\mathcal{C})\eta_j\). This local value inherently exceeds the global curvature \(\mathcal{C}\), explicitly signaling that the ultimate overall measurement precision is not merely a sum of local effects, but relies fundamentally on the presence of metrologically useful inter-ensemble entanglement that coordinates the distributed nodes to further enhance the overall measurement precision.

\section{Connection between QFIM and the multiparameter squeezing matrix}\label{Appendix F}
In this Appendix, we derive the matrix inequality relating the covariance matrices of conjugate quadratures and discuss its connection to state purity and quantum metrological limits. This analysis provides the theoretical basis for the criterion presented in Eq.~\eqref{difference} of the main text.

\subsection{Matrix uncertainty relation}
The Robertson-Schr\"odinger uncertainty principle~\cite{Robertson1929,Weedbrook2012} imposes a constraint on the covariance matrix of any physical state. In our convention, this condition reads
\begin{equation}\label{eq:uncertainty}
	\bm{\Gamma}_Q^a + \upi\bm{\omega} \ge 0,
\end{equation}
where  the symplectic matrix \(\bm{\omega}\) takes the block form due to the ordering of operators,
\begin{equation}
	\bm{\omega} = \begin{pmatrix}
		\mathbf{0} & \mathbbm{1}_M \\
		-\mathbbm{1}_M & \mathbf{0}
	\end{pmatrix}.
\end{equation}
For the steady states considered in this work, the correlations between the \(X\)- and \(Y\)-quadratures vanish (i.e., the off-diagonal blocks of the covariance matrix are zero). Thus, the covariance matrix assumes a block-diagonal form \(\bm{\Gamma}_Q^a = \mathrm{diag}(\bm{\Gamma}_X^a, \bm{\Gamma}_Y^a)\), where \(\bm{\Gamma}_X^a\) and \(\bm{\Gamma}_Y^a\) are the \(M \times M\) covariance matrices for the \(X\) and \(Y\) quadratures, respectively. Substituting this form into Eq.~\eqref{eq:uncertainty}, we obtain
\begin{equation}
	\begin{pmatrix}
		\bm{\Gamma}_X^a & \upi\mathbbm{1}_M \\
		-\upi\mathbbm{1}_M & \bm{\Gamma}_Y^a
	\end{pmatrix} \ge 0.
\end{equation}
Using the Schur complement condition for positive semi-definiteness~\cite{Horn2012}, this inequality holds if and only if
\begin{equation}
	\bm{\Gamma}_X^a - (\upi\mathbbm{1}_M) (\bm{\Gamma}_Y^a)^{-1} (-\upi\mathbbm{1}_M) \ge 0.
\end{equation}
Since \((\upi)(-\upi) = 1\), this simplifies to the elegant matrix inequality:
\begin{equation}\label{eq:matrix_inverse_bound}
	\bm{\Gamma}_X^a \ge (\bm{\Gamma}_Y^a)^{-1}.
\end{equation}
This relation states that the covariance matrix of the \(X\)-quadratures is bounded from below by the inverse of the \(Y\)-quadrature covariance matrix.

\subsection{Gaussian state purity and saturation}
The purity of a Gaussian state in this convention is given by \(p = \mathrm{Tr}(\hat{\rho}^2) = 1/\sqrt{\det(\bm{\Gamma}_Q^a)}\)~\cite{Weedbrook2012,Paris2003}. For a pure state, \(p=1\), which implies \(\det(\bm{\Gamma}_Q^a) = 1\).
In the block-diagonal case, the determinant factorizes as \(\det(\bm{\Gamma}_Q^a) = \det(\bm{\Gamma}_X^a)\det(\bm{\Gamma}_Y^a)\).
When the inequality in Eq.~\eqref{eq:matrix_inverse_bound} is saturated, i.e., \(\bm{\Gamma}_X^a = (\bm{\Gamma}_Y^a)^{-1}\), the determinant becomes
\begin{equation}
	\det(\bm{\Gamma}_Q^a) = \det(\bm{\Gamma}_X^a) \det((\bm{\Gamma}_X^a)^{-1}) = 1.
\end{equation}
Therefore, the equality \(\bm{\Gamma}_X^a - (\bm{\Gamma}_Y^a)^{-1} = 0\) uniquely identifies a pure Gaussian state (a minimum-uncertainty state). Since any deviation from this equality would strictly increase the determinant. Physically, a minimum-uncertainty state refers to a quantum state that strictly saturates the generalized Heisenberg uncertainty relation. For block-diagonal covariance matrices, the uncertainty principle imposes the fundamental matrix bound \(\bm{\Gamma}_X^a \geq (\bm{\Gamma}_Y^a)^{-1}\) in the positive-semidefinite sense. The equality condition \(\bm{\Gamma}_X^a = (\bm{\Gamma}_Y^a)^{-1}\) demonstrates that the quantum fluctuations across all conjugate collective quadratures exactly reach the absolute minimum allowed by quantum mechanics. In the next section, we explicitly demonstrate that the ultimate limit set by the QFIM coincides with the precision bound derived from the multiparameter spin-squeezing matrix.

\subsection{Modified squeezing matrix}
As shown in Eq.~\eqref{SM_MFG}, the moment matrix \(\mathcal{M}_{\boldsymbol{\varphi}}\) is upper bounded by the QFIM \(\bm{F}_Q\). 
In general, obtaining an analytic expression for the QFIM is challenging and often requires numerical methods. For pure probe states, however, this bound can be saturated by \(4\bm{\Gamma}[\hat{\mathbf{G}}]\), which is straightforward to evaluate. Accordingly, the squeezing matrix associated with \(\mathcal{M}_{\boldsymbol{\varphi}}\) admits a corresponding lower bound defined as
\begin{equation}
	\bm{\Xi}^2_{\mathrm{Cov}}[\hat{\mathbf{G}}] = \bm{F}_\mathrm{SN}^{1/2}[\hat{\mathbf{G}}] \left( 4\bm{\Gamma}[\hat{\mathbf{G}}] \right)^{-1} \bm{F}_\mathrm{SN}^{1/2}[\hat{\mathbf{G}}],
\end{equation}
which satisfies
\begin{equation}
	\bm{\Xi}^2 [\hat{\mathbf{G}}, \hat{\mathbf{X}}] \geq \bm{\Xi}^2_{\mathrm{Cov}}[\hat{\mathbf{G}}].
\end{equation}
In our work, we choose \(\hat{\mathbf{G}}=\hat{\mathbf{S}}^Y\), for which the corresponding bound \(\bm{\Xi}^2_\mathrm{Cov}\) coincides with the inverse of the \(Y\)-quadrature block of the dark-state covariance matrix,
\begin{equation}
	\bm{\Xi}^2_\mathrm{Cov}[\hat{\mathbf{S}}^Y] = \left[\bm{\Gamma}_Y^a(\infty)\right]^{-1}.
\end{equation} 
Using the relation established in Eq.~\eqref{SM_Ya_Xa} between \(\bm{\Gamma}_X^a(\infty)\) and \(\left[\bm{\Gamma}_Y^a(\infty)\right]^{-1}\), we may express the squeezing matrix as
\begin{equation}
	\bm{\Xi}^2[\hat{\mathbf{S}}^Y, \hat{\mathbf{S}}^X] = \bm{\Xi}^2_\mathrm{Cov}[\hat{\mathbf{S}}^Y] + \left(\tan\!\phi\right)^2 \mathbf{e}_2 \mathbf{e}_2^\top.
\end{equation} 
For uniform coupling coefficients, the vectors \(\mathbf{A}\) and \(\mathbf{B}\) become parallel, corresponding to \(\phi = 0\). 
In this case,
\begin{equation}
	\bm{\Xi}^2[\hat{\mathbf{S}}^Y, \hat{\mathbf{S}}^X] = \bm{\Xi}^2_\mathrm{Cov}[\hat{\mathbf{S}}^Y].
\end{equation} 
This demonstrates that the squeezing matrix attains its covariance-based lower bound, indicating that the choice of the configuration \(\{\hat{\mathbf{S}}^Y, \hat{\mathbf{S}}^X\}\) is information-theoretically optimal for the pure probe states realized under uniform coupling.


%

\end{document}